\documentclass[aip,prx,reprint]{revtex4-1}

\usepackage{graphicx}
\usepackage{subfigure}
\usepackage{mhchem}

\usepackage{hyperref}

\usepackage{tikz}
\usetikzlibrary{quantikz}

\begin{document}

\title{Chemically-accurate prediction of the ionisation potential of helium using a quantum processor}
\author{Manolo C. Per}
\email{manolo.per@data61.csiro.au}
\affiliation{CSIRO Data61, Clayton, Victoria 3168, Australia}
\author{Nathan Rhodes}
\email{nrhodes@kpmg.com.au}
\affiliation{KPMG Futures, KPMG Australia, Victoria 3008, Australia}
\author{Maiyuren Srikumar}
\affiliation{KPMG Futures, KPMG Australia, New South Wales 2000, Australia}
\author{Joshua W. Dai}
\affiliation{CSIRO Data61, Clayton, Victoria 3168, Australia}

\begin{abstract}
Quantum computers have the potential to revolutionise our understanding of the microscopic behaviour of materials and chemical processes by enabling high-accuracy electronic structure calculations to scale more efficiently than is possible using classical computers.
Current quantum computing hardware devices suffer from the dual challenges of noise and cost, which raises the question of what practical value these devices might offer before full fault tolerance is achieved and economies of scale enable cheaper access. 
Here we examine the practical value of noisy quantum computers as tools for high-accuracy electronic structure, by using a Quantinuum ion-trap quantum computer to predict the ionisation potential of helium.
By combining a series of techniques suited for use with current hardware including qubit-efficient encoding coupled with chemical insight, low-cost variational optimisation with hardware-adapted quantum circuits, and moments-based corrections, we obtain an ionisation potential of 24.5536 $(+0.0011, -0.0005)$ eV, which agrees with the experimentally measured value to within true chemical accuracy, and with high statistical confidence.
The methods employed here can be generalised to predict other properties and expand our understanding of the value that might be provided by near-term quantum computers.
 
\end{abstract}

\date{\today}
\maketitle

\section{Introduction}

First-principles electronic structure methods are powerful tools for elucidating the microscopic properties of materials and chemical processes.  
As the cost of performing these calculations is extremely high, in practice they are limited to investigating relatively small systems.
Approximations, such as those based on Density Functional Theory \cite{Teale2024}, are less computationally demanding, and are widely used to provide qualitative insight.
However, these methods have some serious limitations when accurate quantitative predictions are required for practical challenges such as heterogeneous catalysis \cite{Oudot2024} and biomolecules containing transition metals \cite{Radon2024}.

Quantum computing holds great promise as a tool for expanding the scope of electronic structure calculations.
One of the original motivations for building a quantum computer was Feynman's observation \cite{Feynman1981} that such a device would be essential for efficiently simulating the quantum behaviour of Nature.
While there are known quantum algorithms \cite{Abrams1999} which have the potential to deliver revolutionary improvements over the best known classical-computing methods, it is expected that full fault tolerance will be required to implement them reliably, which is far beyond the capabilities of current quantum computers \cite{Blunt2024}. 
Recently, there has been progress in demonstrating the advantage of logical qubits compared to physical qubits for a chemical system using simpler algorithms \cite{vandam2024}.

In the meantime until full fault tolerance is achieved, there is an important question as to what practical value noisy quantum computers can provide.
In particular, as the expected value in applying quantum computing to electronic structure problems is to provide the high quantitative accuracy that is challenging for classical computing, the primary question is whether noisy quantum computers can provide sufficiently accurate results.
The transcorrelated approach of Dobrautz \textit{et al.} \cite{Dobrautz2024} shows promise as one way of reducing the number of qubits required to reach high accuracy, but the precision of results obtained on superconducting hardware suffers from large statistical uncertainties due to the practical cost of performing error mitigated calculations.

In this work, we explore the challenge of practically achieving high accuracy and precision by using a Quantinuum ion-trap quantum processor to predict the ionisation potential  of helium.
The ionisation potential (IP) of Helium is known to high accuracy experimentally, with a value of 24.58737618(2) eV obtained using extreme ultraviolet frequency comb metrology \cite{Kandula2010}.
Computationally, the IP can be calculated as the difference in energy between the neutral He atom and the \ce{He+} cation. (Fig.~\ref{fig:map} (a)).
As the cation is a single-electron system, its energy can be calculated trivially, and so in this work we focus on using a quantum computer to accurately calculate the energy of the neutral \ce{He} atom.

The rest of this paper is organised as follows.
In Section~\ref{sec:rep} we describe our approach to mapping the problem onto the quantum computer, including resource reduction techniques.
In Section~\ref{sec:vqe} we detail the quantum algorithm and the methods used to optimise circuit parameters.
In Section~\ref{sec:ip} we present the hardware experiments used to calculate the IP.

\section{Representation}\label{sec:rep}

\begin{figure*}
\begin{center}
\resizebox*{16cm}{!}{\includegraphics{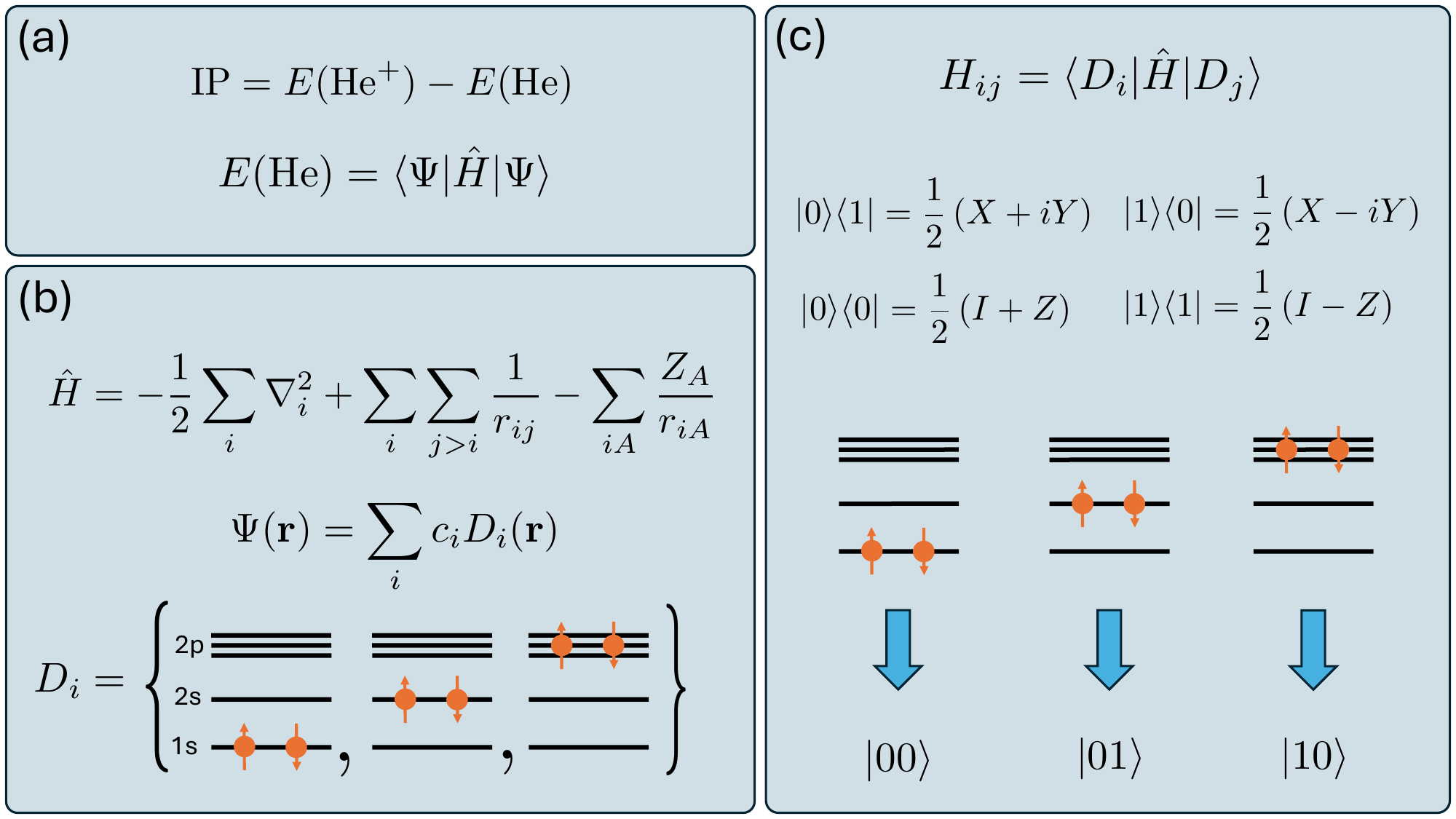}}
\caption{\textbf{Mapping the physical problem onto the quantum computer.}
(a). The ionisation potential is defined as the total energy difference between the cation and neutral atom. Only the energy of the neutral helium atom is required, as the single-electron cation has a known closed-form solution.
(b). We use a first-quantised approach in which the wavefunction is expanded as a sum of real-space many-body basis functions with unknown weights $c_{i}$. The number of basis functions is reduced by restricting the terms to seniority-zero excitations from the Hartree-Fock reference, and imposition of spherical symmetry.
(c). Bit-wise binary encoding of the Hamiltonian matrix generates a qubit-Hamiltonian and maps the many-body basis states onto the quantum computational basis states.}
\label{fig:map}
\end{center}
\end{figure*}

Our first step is to decide how to represent the physical problem we are tying to solve on a quantum computer.
The most commonly used approach for representing electronic structure problems on quantum computers is to use the second-quantised formalism \cite{Cao2019}, in which the electronic Hamiltonian is written in terms of electron creation and annihilation operators. 
While this approach has benefits, including a number of terms that scales polynomially with the system size, it has a number of drawbacks for performing high-accuracy calculations on small, noisy quantum devices.
In the second-quantised representation, the number of qubits required is comparable to the number of electronic spin-orbitals (the exact number of qubits depends on the details of the chosen mapping procedure \cite{Setia2020}).
The number of orbitals is related to basis set size, which is a measure of the coarseness of the discretisation of space, so highly accurate calculations require large basis sets, and hence large numbers of qubits (see Appendix~\ref{app:basis}).
Large numbers of qubits require commensurately large numbers of quantum operations to implement algorithms, which leads to problems with noise.
An additional drawback of the second-quantised representation is that it maps the entire Fock space, so the number of unphysical states (i.e. those containing a number of electrons other than the one we are interested in) increases exponentially with system size \cite{Babbush2018}.
This can further reduce accuracy, as noisy quantum devices can fall into unphysical states even if the quantum algorithm used is designed to avoid them. \cite{Bonet-Monroig2018}
As a result of these drawbacks, the vast majority of electronic structure demonstrations on quantum computers using second-quantisation have been performed with small, usually minimal, single-particle basis sets that cannot provide highly accurate descriptions of the physical system being investigated.

Here, we instead use a first-quantised approach, starting from the real-space Schr\"odinger Hamiltonian, and a wavefunction expanded as a linear combination of Slater determinants \cite{Szabo} (Fig.~\ref{fig:map}(b)).
While first-quantised approaches have received less attention than second-quantised approaches for quantum computed chemistry, they have the potential to enable high accuracy calculations in a qubit-efficient way \cite{Volkmann2024}.
The single-particle atom-centered basis that we use admits 25 Slater determinants, including single and double excitations from the Hartree-Fock reference state.
We reduce the size of the determinant basis by allowing only paired double excitations from the Hartree-Fock state - the so-called seniority-zero space.
This is known to be the dominant sector of the space of determinants, and has been used before in second-quantised electronic-structure calculations on quantum computers \cite{Elfving2021,Zhao2023}.

We further reduce the number of Slater determinants in our wavefunction by enforcing the spherical symmetry of the $^1S$ ground-state of Helium.
We do this by constructing a linear combination of the Slater determinants corresponding to paired occupancies of the atomic $2p_{x},2p_{y},2p_{z}$ states (known in the quantum chemistry literature as a Configuration State Function).

Together, these simplifications result in three many-body basis functions in which to expand our wavefunction. (Fig.\ref{fig:map}(b)).
The resulting matrix Hamiltonian $H_{ij}$, involving matrix elements between these many-body basis functions, is known as a selected configuration-interaction Hamiltonian. \cite{Szabo}
It can be written in a form suitable for representation on a quantum computer by using binary encoding \cite{Sawaya2020} (Fig.~\ref{fig:map}(c)), which produces a linear combination of Pauli strings.
This approach is completely general and enables $N_{\textrm{det}}$ determinants to be mapped onto $\log_{2} (N_{\textrm{det}})$ qubits.
As we have a $3\times 3$ Hamiltonian matrix, we can map it onto 2 qubits.
The binary encoding maps each many-body basis state onto a unique computational basis state, with one of the four computational basis states unused.
We construct the mapping to take advantage of the fact that the qubit ground-state is likely to be less noisy than excited states, by mapping the Hartree-Fock determinant (which we expect to dominate) onto the $|00\rangle$ state, and leaving the $|11\rangle$ basis state unused.

With this binary encoding, our qubit Hamiltonian contains 10 terms:
\begin{equation}\label{eq:ham}
\begin{split}
H =&  -0.0336 \;\textrm{II} -0.4872 \;\textrm{IZ} -0.9436 \;\textrm{ZI}  -1.3971 \;\textrm{ZZ} \\ 
& + 0.1239\;\textrm{IX} + 0.12392\;\textrm{ZX} + 0.1548\;\textrm{XI} \\
& + 0.1548\;\textrm{XZ} + 0.0379\;\textrm{XX} + 0.0379\;\textrm{YY}
\end{split}
\end{equation}
We determine the ground-state energy of this Hamiltonian using hybrid quantum algorithms which are suited to implementation on current noisy quantum computers.
These involve using the quantum computer to calculate expectation values of the terms in the qubit Hamiltonian.
The cost of evaluating these expectation values can be reduced significantly by taking advantage of the fact that many of the terms in the qubit Hamiltonian qubit-wise commute, and so can be measured simultaneously.
Only 5 independent basis measurements (YY, ZZ, ZX, XZ, XX) are required.

\section{Variational Optimisation}\label{sec:vqe}

\begin{figure*}
\centering
\subfigure[]{
\resizebox*{5.8cm}{!}{ 
\begin{quantikz} 
\lstick{\ket{0}} & \qw & \gate{R_{y}(\theta_{2})} & \qw & \qw \\
\lstick{\ket{0}} & \gate{R_y(\theta_{1})} & \ctrl{-1} & \gate{X} & \qw
\end{quantikz}
}
}
\subfigure[]{
\resizebox*{10cm}{!}{
\begin{quantikz} 
\lstick{\ket{0}} & \gate{S^{\dagger}} & \gate{H} & \gate{R_{z}(\theta_{2}/2)} & \gate[wires=2]{R_{ZZ}(-\theta_{2}/2)}& \gate{H} & \gate{S} \\
\lstick{\ket{0}} & \gate{R_y(\theta_{1})} & \qw & \qw & \qw & \gate{X} & \qw
\end{quantikz}
}
}
\subfigure[]{
\resizebox*{8cm}{!}{\includegraphics{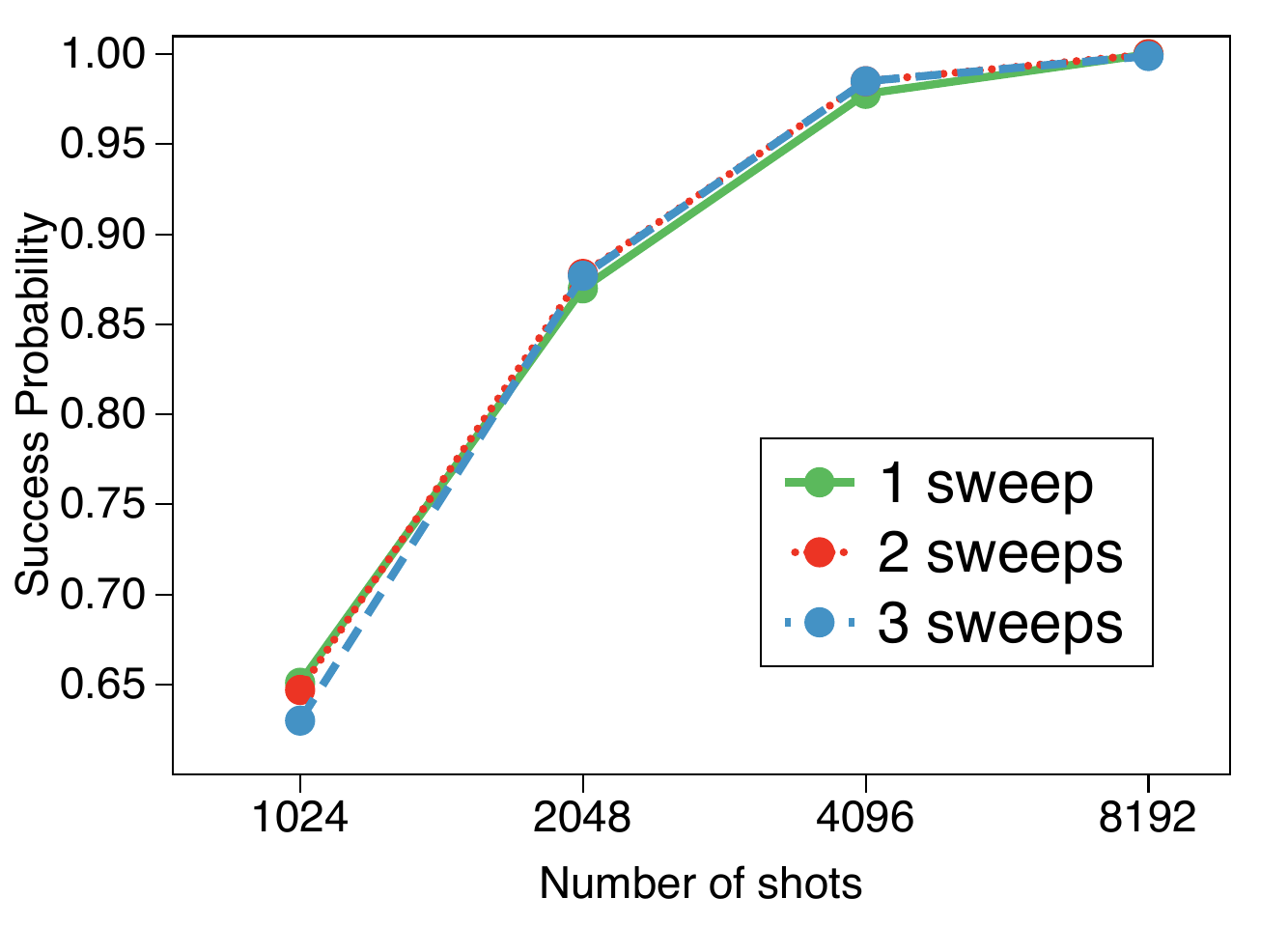}}}
\subfigure[]{
\resizebox*{8cm}{!}{\includegraphics{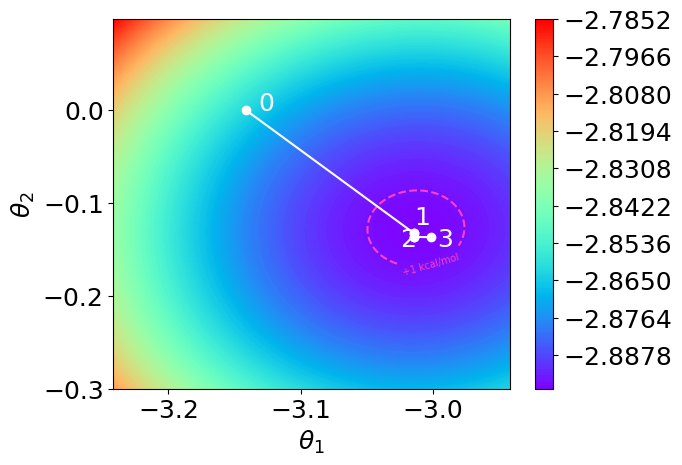}}}\\
\caption{\label{fig:vqe} \textbf{VQE calculations.} (a) Original form of the parameterised quantum circuit encoding the variational wavefunction ansatz. (b) Re-factoring of the quantum circuit into gates native to the Quantinuum H1-1 device. Note the circuit is further optimised in Appendix~\ref{app:circuits}. (c) Results of simulations of the optimisation process, showing that small numbers of shots are sufficient to have a high probability of obtaining accurate variational parameters. (d) Results of the full VQE process performed on the Quantinuum H1-1 Emulator. The white points indicate the parameter values at each sweep, starting from the Hartree-Fock values. The pink dashed line indicates the region inside which the parameters lead to $\langle H \rangle$ within 1 kcal/mol of the exact (in-basis) energy. The background colour gradient indicates the total energy (in Hartree) corresponding to the parameters, in the absence of device noise. }
\end{figure*}

The Variational Quantum Eigensolver (VQE) \cite{Peruzzo2014} is a practical method for performing electronic structure calculations on noisy quantum computers.
It is a hybrid approach that uses the quantum computer as a co-processor to efficiently evaluate expectation values of the Hamiltonian in some user-defined parameterised quantum state.
Using the fact that these expectation values are an upper bound to the ground-state energy, $\langle H \rangle \geq E_{0}$,  a classical computer can be used to find the parameters that minimise the expectation value and thus provide the most accurate estimate of the ground-state energy.
Two key components of practical VQE calculations are the form of the quantum circuit that prepares the parameterised quantum state, and the choice of optimisation method.

The basic form of the quantum circuit we use for state preparation is shown in Fig.~\ref{fig:vqe}(a).
It is straightforward to verify that this circuit can produce any superposition of the three computational basis states encoding our wavefunction in Fig.~\ref{fig:map}(d), and avoids the unphysical $|11\rangle$ basis state, when initialised in the Hartree-Fock ($|00\rangle$) state.
In order to reduce the impact of noise as much as possible, we re-factor this basic circuit in terms of the parameterised $R_{zz}$ gate that is native to the Quantinuum H1-1 device (Fig.~\ref{fig:vqe}(b)).
We evaluate expectation values as projective measurements in the computational basis with appropriate single-qubit post-rotations for each of the five independent measurement bases. (see Appendix~\ref{app:circuits} for details of the actual circuits implemented, including simplifications made for efficiency).

The use of a finite number of shots for performing the measurements introduces an inherent statistical uncertainty into the expectation values, independent of any device noise.
It is important to use an optimisation method in VQE that can efficiently cope with this uncertainty.
Here we use a gradient-free approach that makes use of the underlying trigonometric structure of the parameterised quantum gates, based on the approach in Ref.~\onlinecite{Ostaszewski2021} (see Appendix~\ref{app:opt}).
We use a greedy approach, in which each parameter is individually optimised while holding the other parameter fixed, and each pass over both parameters is termed a sweep.

Because of the need for multiple evaluations of the expectation value of the Hamiltonian, the VQE process can be extremely expensive.
 Critical hyperparameters that determine both the cost and accuracy of the process are the number of sweeps and the number of shots used for each expectation value.
The number of shots affects the amount of statistical noise in the individual estimates of the expectation values, and this noise can hinder the ability of the optimisation method to accurately find the optimal variational parameters.
As a result, the cost of accessing sufficient quantum resources to converge the variational parameters using VQE can become a limiting factor in practical applications.
In Sec.~\ref{sec:ip} we implement a method which has shown promise in helping to overcome this issue.

In order to investigate the behaviour of the hyperparameters for our application, we use a probabilistic measure of a successful optimisation.
We define this to be the probability that a set of variational parameters $(\theta_{1},\theta_{2})$ obtained from VQE produce a value of $\langle H \rangle$ which is within 1 kcal/mol of the exact ground-state energy (in the basis used to represent the wavefunction).
We calculate this success probability as a function of the number of shots and sweeps by performing repeated idealised simulations (i.e. without any device noise), and simply counting the fraction of simulations that result in sufficiently accurate parameters.
Remarkably, the optimisation method we use is relatively insensitive to both the number of sweeps and the noise in the expectation values (Fig.~\ref{fig:vqe}(c)).
We obtain $>95\%$ chance of obtaining accurate parameters with only 4,096 shots per expectation value, which is orders of magnitude fewer than that required to evaluate $\langle H \rangle$ itself to similar statistical accuracy (see Fig.3(d)).
We note that these conclusions were obtained using idealised simulations which did not include the effect of quantum device noise.
To account for this, in our practical calculations we were less frugal and used 3 sweeps over the parameters, with 8,192 shots per expectation value. 
These calculations were performed on the Quantinuum H1-1 emulator, in order to save our hardware budget for the final calculation of the energy.
We note that based on our calculations, the H1-1 emulator produces expectation values very similar to the quantum hardware, so we expect that similar results would be obtained from end-to-end hardware VQE calculations, given a sufficiently large hardware shot budget.

The VQE optimisation procedure on the H1-1 emulator was started from parameter values of $(-\pi,0)$. These correspond to the Hartree-Fock energy, in that they leave the initial $|00\rangle$ quantum state unchanged, and are expected to be a reasonable starting point for a single-reference system.
As expected from our idealised simulations, the parameter values converged quickly, and a single optimisation sweep was sufficient to produce accurate parameters (Fig.~\ref{fig:vqe}(d)).
The final parameters obtained after three sweeps are $(-3.0016,-0.1370)$.
Note that the energies shown in Fig.~\ref{fig:vqe}(d) are the \textit{exact} in-basis values of $\langle H \rangle$ corresponding to the parameters $(\theta_{1},\theta_{2})$, and \textit{not} those obtained on the H1-1 emulator or real device.
As will be shown later, the values of $\langle H \rangle$ calculated on the real device deviate significantly from these exact values.
The fact that the VQE process produced such accurate parameters suggests that device noise has shifted the energy by a roughly constant amount, retaining its relative shape in the vicinity of the minimum. 
A similar effect was observed in Ref.~\onlinecite{Zhao2023}.

\section{Hardware Experiments}\label{sec:ip}

\begin{figure*}
\begin{center}
\subfigure[]{
\resizebox*{8cm}{!}{\includegraphics{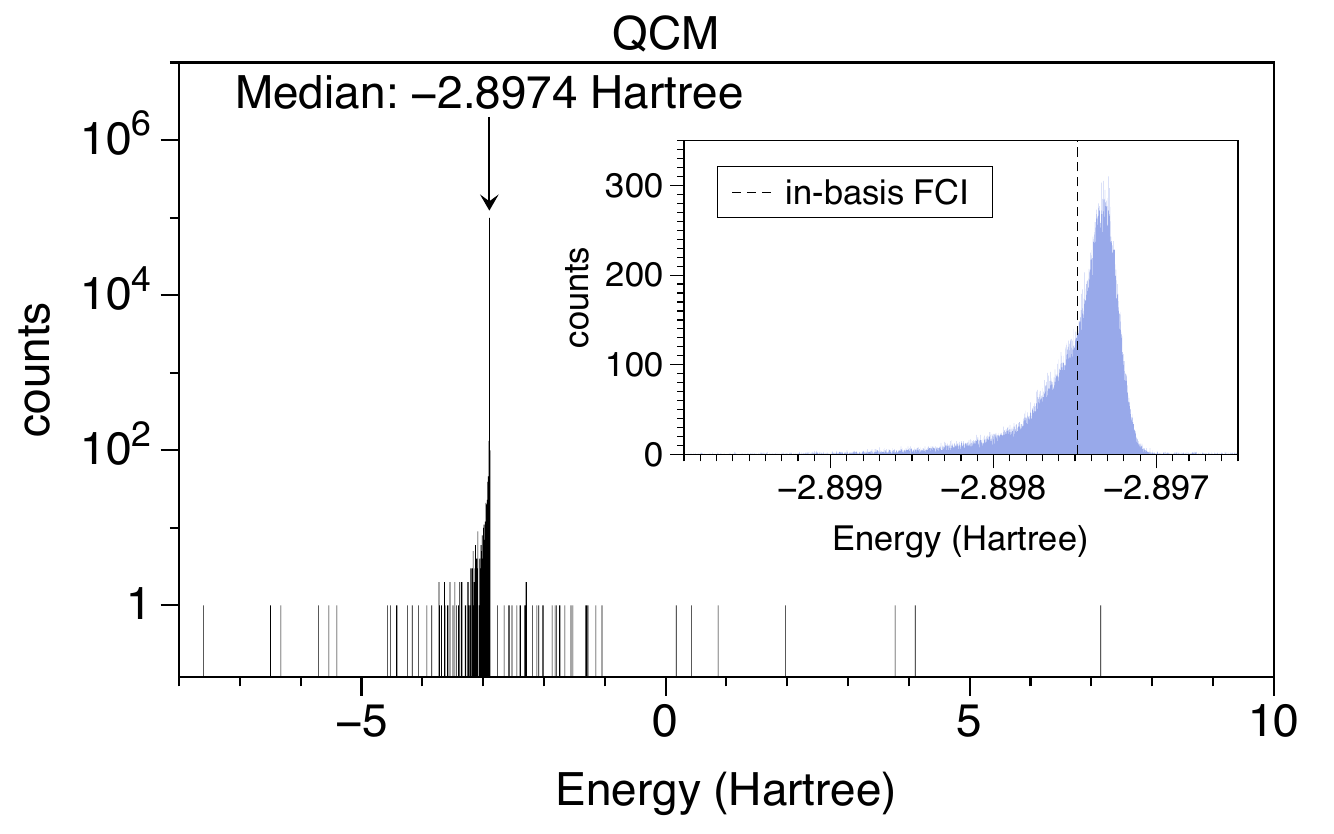}}}
\subfigure[]{
\resizebox*{8cm}{!}{\includegraphics{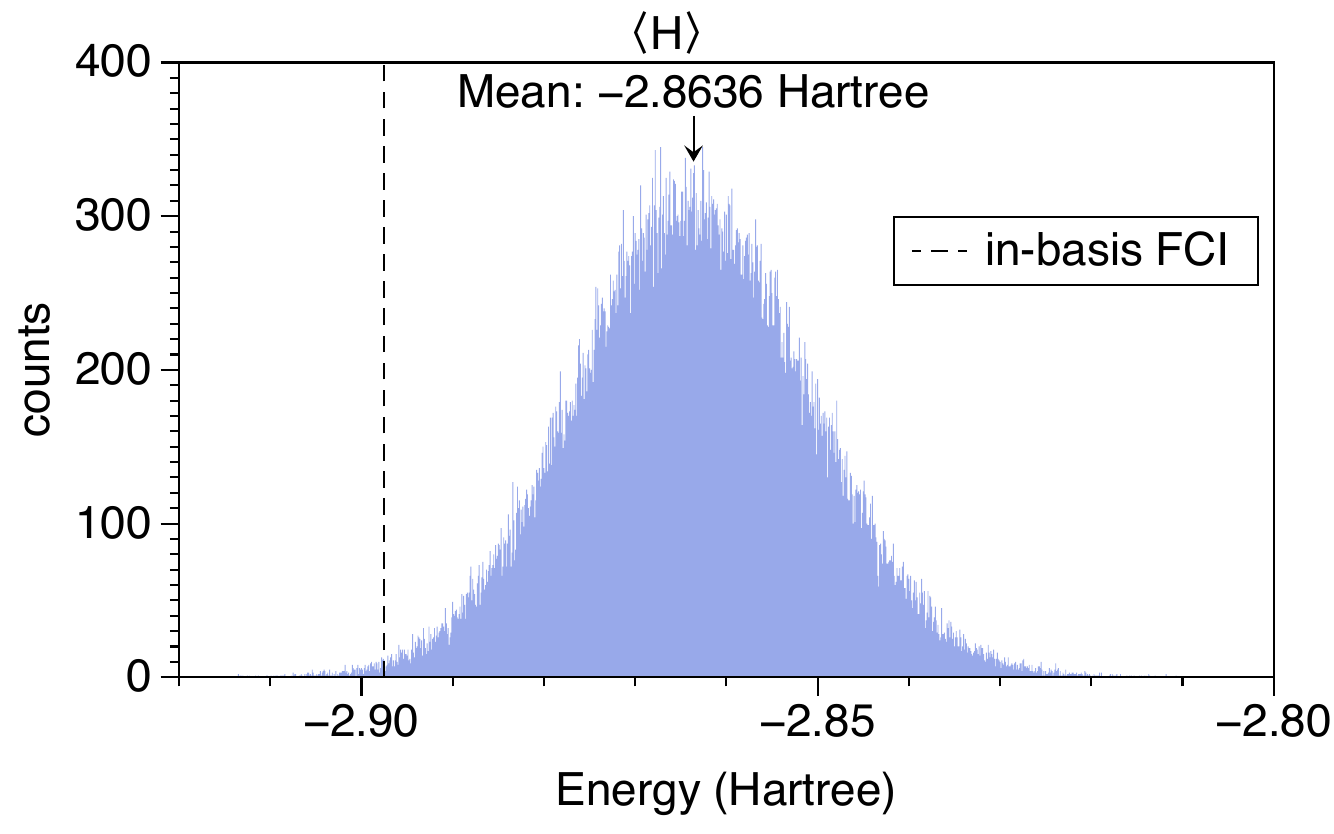}}}\\
\subfigure[]{
\resizebox*{8cm}{!}{\includegraphics{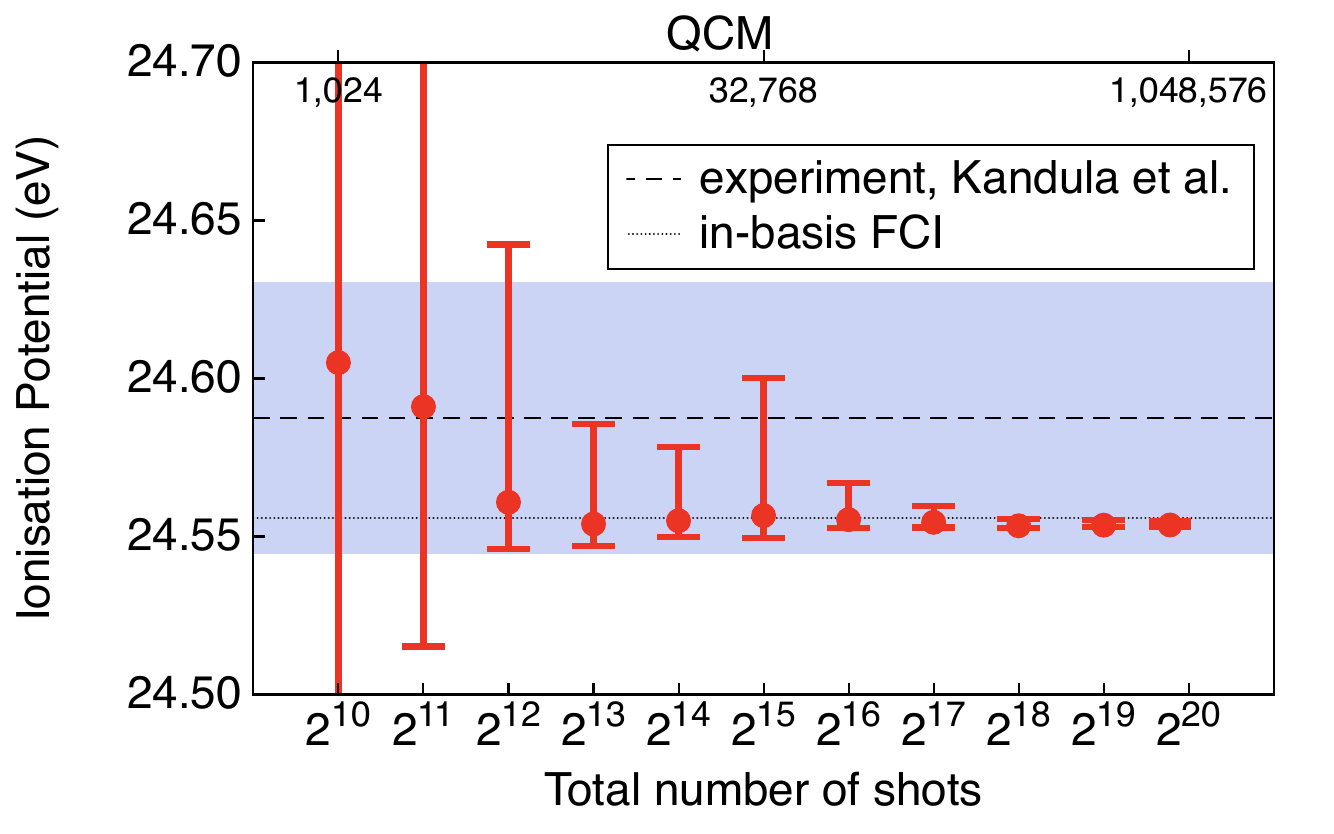}}}
\subfigure[]{
\resizebox*{8cm}{!}{\includegraphics{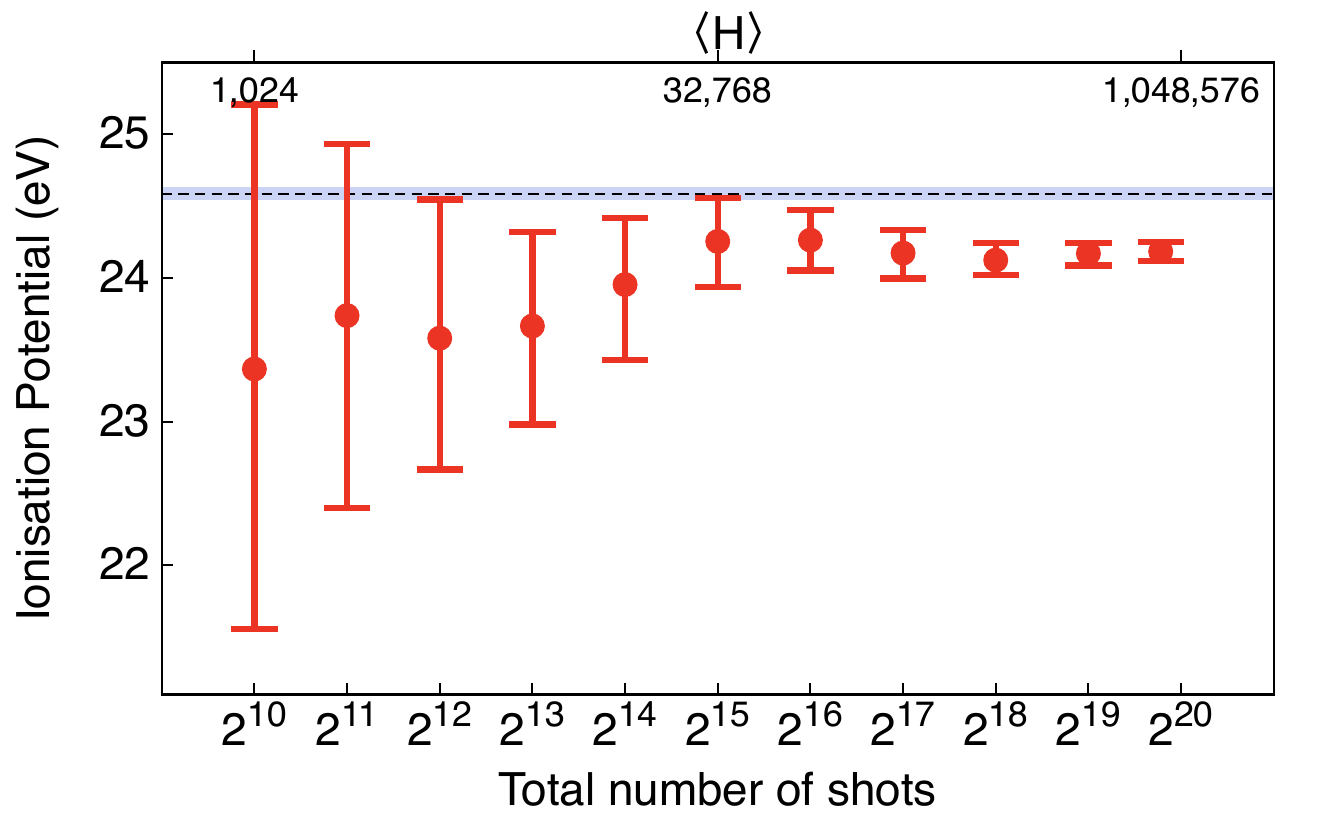}}}\\
\caption{\label{fig:ip} \textbf{Results of hardware experiments performed on the Quantinuum H1-1 device.} (a) Distribution of bootstrapped QCM energies of the He atom, showing the presence of extreme outliers (note the logarithmic vertical scale), and the asymmetric shape of the distribution (inset). (b) Distribution of bootstrapped values of $\langle H \rangle$ for the He atom, showing a normal distribution. (c) Ionisation potential of He obtained using QCM energies, as a function of the total number of shots used in the quantum measurements. (d) Ionisation potential of He obtained using $\langle H \rangle$, as a function of the total number of shots used in the quantum measurements. In both (c) and (d), the dashed horizontal line marks the experimental value of $24.58737618$ eV from Ref.~\onlinecite{Kandula2010}, while the light blue band indicates the region of chemical accuracy, $\pm 0.043$ eV (1 kcal/mol) about the experimental value. Error bars indicate the extent of the central $95^{\textrm{th}}$ percentile (equivalent to approximately $\pm 2\sigma$ for a normal distribution).}
\end{center}
\end{figure*}

Now that we have obtained the optimal parameters for the variational ansatz circuit, we use them to calculate the final energy, and hence IP, using the Quantinuum H1-1 quantum device.

To do this, we use the Quantum Computed Moments (QCM) method. \cite{Vallury2020}
QCM is a non-variational approach based on a Lanczos cumulant expansion \cite{Hollenberg1993}, using expectation values of powers of the Hamiltonian $\langle H^{n} \rangle$ up to $n=4$ (though the method can be generalised to use higher powers \cite{Witte1997}).
Using quantum-computed expectation values of the powers of the Hamiltonian, cumulants are constructed,
\begin{equation}
c_{n} = \langle H^{n} \rangle - \sum_{p=0}^{n-2} {n-1 \choose p} c_{p+1} \langle H^{n-1-p }\rangle	
\end{equation}
and these cumulants are then used in the closed-form expression due to Hollenberg and Witte (HW) \cite{Hollenberg1994},
\begin{equation}
E_{\textrm{HW}} = c_{1} - \frac{c_{2}}{c_{3}^{2}-c_{2}c_{4}} \left[ \sqrt{3c_{3}^{2} -2 c_{2} c_{4}} - c_{3} \right]
\end{equation}
The QCM approach has recently been successfully applied to a number of applications in quantum computing, including minimal-basis second-quantised electronic-structure demonstrations \cite{Jones2022}, where it has been shown to be accurate and highly robust in the presence of device noise for ground-state energy estimation \cite{Jones2024}.
QCM has also recently been used to calculate the single particle Green's function in a spin-orbital basis. \cite{GreeneDiniz2024}

In previous second-quantised applications, the calculation of $\langle H^{n}\rangle$ for $n>1$ has required additional quantum measurements over those required for $n=1$.
This is because the encoding of the powers of the Hamiltonian as Pauli strings leads to terms not present in the description of the original Hamiltonian.
With the dense binary encoding used in this work (Fig.~\ref{fig:map}(c)), powers of the Hamiltonian matrix can be represented using the same list of Pauli strings as the original Hamiltonian (only the coefficients of the strings will differ).
Therefore the expectation values of the required powers $\langle H^{n}\rangle$ for $n=2,3,4$ can be evaluated using the same set of quantum measurements as required for the Hamiltonian itself ($n=1$).
Note that efficiently generalising the methods implemented here to much larger systems will require making use of the sparsity properties of the configuration interaction matrix Hamiltonian \cite{Toloui2013,Babbush2018,Kirby2021} and in this case we expect that expectation values of powers of the sparse Hamiltonian will require additional quantum measurements over those required for $\langle H \rangle$.

Using the variational parameters obtained from the previous VQE calculations, we used the Quantinuum H1-1 device to calculate the expectation values of each of the 5 measurement bases.
In order to maximise efficiency, the total shot budget was apportioned between the bases in such a way as to minimise the variance of the total energy (see Appendix~\ref{app:shot}), and thus obtain the smallest statistical uncertainty possible for a given number of shots.
As with the calculations on the emulator, the expectation values were calculated by repeatedly preparing the quantum state circuits and making measurements in the computational basis.
In total 776,900 hardware shots were performed across 34 sessions on the device.
Each session contained one job for each measurement basis, for a total of 170 circuits.

Bootstrapping was used to perform statistical analyses, by re-sampling with replacement at the level of the individual shot results (i.e. the computational basis state obtained, $\{00,01,10,11\}$), and using the re-sampled list of results to calculate expectation values and QCM energies.
The distribution of QCM energies obtained by this bootstrapping is shown in Fig.~\ref{fig:ip}(a).
The bulk portion of the distribution (shown in the inset figure) is asymmetric with a long negative tail, but extreme outliers occur on both sides.
As a result, the spread of the distribution cannot be characterised using a mean and variance, so we describe the spread of the QCM using the median and middle $95^{\textrm{th}}$ percentile. 
In contrast, the bootstrapped distribution of $\langle H\rangle$ shown in Fig.~\ref{fig:ip}(b) has the familiar normal form.

The IP of He was constructed from the QCM energy and from $\langle H \rangle$, both using expectation values obtained from the physical device.
In Fig.~\ref{fig:ip}(c,d), these estimates of the IP are shown as functions of the number of shots used, in order to examine the convergence of the IP and its uncertainty for both methods.
For each number of total shots $N$, the bootstrapping analysis was performed using only the results of the first $N$ shots, so each point is equivalent to having performed only $N$ total shots, and contains no additional information from the larger dataset.

The QCM calculations produce highly accurate IPs.
Using just $2^{13} =$ 8,192 shots, we obtain an IP that is within true chemical accuracy of the experimental value, with $95\%$ confidence. (Fig.~\ref{fig:ip}(c)).
The asymmetry of the distribution is evident in the error bars, with the long negative energy tail translating into greater uncertainty in the direction of increasing IP.
As the number of shots is increased, the QCM IPs converge to the exact in-basis result.
Note that the small difference between the exact in-basis IP and the experimental IP is due to the small single-particle basis set used to build the orbitals in our many-body basis functions.
Obtaining results even closer to experiment would require the use of a larger single-particle basis set, and likely also more determinants.

It is illuminating to compare the QCM IP results with those constructed using the bare $\langle H \rangle$ values, shown in Fig.~\ref{fig:ip}(d).
Using the same number of shots as QCM, the raw expectation value of the Hamiltonian results in IPs which are far from chemical accuracy, converging to a value which deviates from the experimental result by -0.404 eV (-9.3 kcal/mol).
These results also display much larger statistical error bars than QCM for the same number of shots.
With the largest number of shots used (776,900) the middle $95^{\textrm{th}}$ percentile of the bootstrapped distribution spans a range of 0.13 eV for the raw expectation value, but for QCM this range is nearly two orders smaller, at 0.0016 eV.
We note that implementation of further error mitigation might improve the accuracy of the IPs obtained using $\langle H \rangle$, but mitigation techniques typically come at the cost of requiring more shots \cite{Temme2017} or qubits \cite{Huggins2021}, making error mitigated calculations even more expensive.

\section{Conclusion}

This work is a successful attempt at using a noisy quantum computer to predict an observable property of Nature in close quantitative agreement with experiment.
Our best result for the ionisation potential of He is 24.5536 $(+0.0011, -0.0005)$ eV, which deviates from the experimental value by only 0.034 eV (0.78 kcal/mol), and thus lies within true chemical accuracy to high confidence.

Mapping the space of problems for which close agreement with experiment can be obtained provides a way of charting the progress of quantum computers as practical tools in quantum chemistry and materials science.
The fusion of ideas from traditional electronic-structure theory and quantum computing provides a promising way forward in near-term quantum computational chemistry.
We anticipate that refinements of the techniques implemented here will enable the accurate prediction of more physically measurable quantities, expanding the value of near-term noisy quantum computers.

\section{Acknowledgements}

We would like to thank Michael Egan from KPMG Australia and Chris Vale from CSIRO for their continued support throughout the project. 
Theoretical work was supported by the CSIRO Quantum Technologies Future Science Platform. 
Access to quantum hardware and circuit implementation expertise was provided by KPMG Australia.
J.W.D. was supported by a CSIRO Undergraduate Vacation Scholarship. 
We would also like to thank Quantinuum and Microsoft for valuable discussions on device use and access.

\section{Author Contributions}
M.C.P. and J.W.D. developed the mapping of the physical problem to quantum states. 
M.C.P. implemented mapping of the problem, developed the optimiser for 1 and 2 qubit operators, implemented the QCM method and analysed the final results. 
N.R. developed the hardware-optimised native circuit, implemented the VQE circuit on the ion trap device and extracted data for analysis. 
M.S. determined parameter domain for VQE and optimal shot configurations. 
All authors contributed to writing the manuscript.

\bibliography{refs}

\begin{thebibliography}{46}%
\makeatletter
\providecommand \@ifxundefined [1]{%
 \@ifx{#1\undefined}
}%
\providecommand \@ifnum [1]{%
 \ifnum #1\expandafter \@firstoftwo
 \else \expandafter \@secondoftwo
 \fi
}%
\providecommand \@ifx [1]{%
 \ifx #1\expandafter \@firstoftwo
 \else \expandafter \@secondoftwo
 \fi
}%
\providecommand \natexlab [1]{#1}%
\providecommand \enquote  [1]{``#1''}%
\providecommand \bibnamefont  [1]{#1}%
\providecommand \bibfnamefont [1]{#1}%
\providecommand \citenamefont [1]{#1}%
\providecommand \href@noop [0]{\@secondoftwo}%
\providecommand \href [0]{\begingroup \@sanitize@url \@href}%
\providecommand \@href[1]{\@@startlink{#1}\@@href}%
\providecommand \@@href[1]{\endgroup#1\@@endlink}%
\providecommand \@sanitize@url [0]{\catcode `\\12\catcode `\$12\catcode `\&12\catcode `\#12\catcode `\^12\catcode `\_12\catcode `\%12\relax}%
\providecommand \@@startlink[1]{}%
\providecommand \@@endlink[0]{}%
\providecommand \url  [0]{\begingroup\@sanitize@url \@url }%
\providecommand \@url [1]{\endgroup\@href {#1}{\urlprefix }}%
\providecommand \urlprefix  [0]{URL }%
\providecommand \Eprint [0]{\href }%
\providecommand \doibase [0]{http://dx.doi.org/}%
\providecommand \selectlanguage [0]{\@gobble}%
\providecommand \bibinfo  [0]{\@secondoftwo}%
\providecommand \bibfield  [0]{\@secondoftwo}%
\providecommand \translation [1]{[#1]}%
\providecommand \BibitemOpen [0]{}%
\providecommand \bibitemStop [0]{}%
\providecommand \bibitemNoStop [0]{.\EOS\space}%
\providecommand \EOS [0]{\spacefactor3000\relax}%
\providecommand \BibitemShut  [1]{\csname bibitem#1\endcsname}%
\let\auto@bib@innerbib\@empty
\bibitem [{\citenamefont {Teale}\ \emph {et~al.}(2022)\citenamefont {Teale}, \citenamefont {Helgaker}, \citenamefont {Savin}, \citenamefont {Adamo}, \citenamefont {Aradi}, \citenamefont {Arbuznikov}, \citenamefont {Ayers}, \citenamefont {Baerends}, \citenamefont {Barone}, \citenamefont {Calaminici}, \citenamefont {Canc{\`e}s}, \citenamefont {Carter}, \citenamefont {Chattaraj}, \citenamefont {Chermette}, \citenamefont {Ciofini}, \citenamefont {Crawford}, \citenamefont {De~Proft}, \citenamefont {Dobson}, \citenamefont {Draxl}, \citenamefont {Frauenheim}, \citenamefont {Fromager}, \citenamefont {Fuentealba}, \citenamefont {Gagliardi}, \citenamefont {Galli}, \citenamefont {Gao}, \citenamefont {Geerlings}, \citenamefont {Gidopoulos}, \citenamefont {Gill}, \citenamefont {Gori-Giorgi}, \citenamefont {G{\"o}rling}, \citenamefont {Gould}, \citenamefont {Grimme}, \citenamefont {Gritsenko}, \citenamefont {Jensen}, \citenamefont {Johnson}, \citenamefont {Jones}, \citenamefont {Kaupp}, \citenamefont {K{\"o}ster}, \citenamefont {Kronik}, \citenamefont {Krylov}, \citenamefont {Kvaal}, \citenamefont {Laestadius}, \citenamefont {Levy}, \citenamefont {Lewin}, \citenamefont {Liu}, \citenamefont {Loos}, \citenamefont {Maitra}, \citenamefont {Neese}, \citenamefont {Perdew}, \citenamefont {Pernal}, \citenamefont {Pernot}, \citenamefont {Piecuch}, \citenamefont {Rebolini}, \citenamefont {Reining}, \citenamefont {Romaniello}, \citenamefont {Ruzsinszky}, \citenamefont {Salahub}, \citenamefont {Scheffler}, \citenamefont {Schwerdtfeger}, \citenamefont {Staroverov}, \citenamefont {Sun}, \citenamefont {Tellgren}, \citenamefont {Tozer}, \citenamefont {Trickey}, \citenamefont {Ullrich}, \citenamefont {Vela}, \citenamefont {Vignale}, \citenamefont {Wesolowski}, \citenamefont {Xu},\ and\ \citenamefont {Yang}}]{Teale2024}%
  \BibitemOpen
  \bibfield  {author} {\bibinfo {author} {\bibfnamefont {A.~M.}\ \bibnamefont {Teale}}, \bibinfo {author} {\bibfnamefont {T.}~\bibnamefont {Helgaker}}, \bibinfo {author} {\bibfnamefont {A.}~\bibnamefont {Savin}}, \bibinfo {author} {\bibfnamefont {C.}~\bibnamefont {Adamo}}, \bibinfo {author} {\bibfnamefont {B.}~\bibnamefont {Aradi}}, \bibinfo {author} {\bibfnamefont {A.~V.}\ \bibnamefont {Arbuznikov}}, \bibinfo {author} {\bibfnamefont {P.~W.}\ \bibnamefont {Ayers}}, \bibinfo {author} {\bibfnamefont {E.~J.}\ \bibnamefont {Baerends}}, \bibinfo {author} {\bibfnamefont {V.}~\bibnamefont {Barone}}, \bibinfo {author} {\bibfnamefont {P.}~\bibnamefont {Calaminici}}, \bibinfo {author} {\bibfnamefont {E.}~\bibnamefont {Canc{\`e}s}}, \bibinfo {author} {\bibfnamefont {E.~A.}\ \bibnamefont {Carter}}, \bibinfo {author} {\bibfnamefont {P.~K.}\ \bibnamefont {Chattaraj}}, \bibinfo {author} {\bibfnamefont {H.}~\bibnamefont {Chermette}}, \bibinfo {author} {\bibfnamefont {I.}~\bibnamefont {Ciofini}}, \bibinfo {author} {\bibfnamefont {T.~D.}\ \bibnamefont {Crawford}}, \bibinfo {author} {\bibfnamefont {F.}~\bibnamefont {De~Proft}}, \bibinfo {author} {\bibfnamefont {J.~F.}\ \bibnamefont {Dobson}}, \bibinfo {author} {\bibfnamefont {C.}~\bibnamefont {Draxl}}, \bibinfo {author} {\bibfnamefont {T.}~\bibnamefont {Frauenheim}}, \bibinfo {author} {\bibfnamefont {E.}~\bibnamefont {Fromager}}, \bibinfo {author} {\bibfnamefont {P.}~\bibnamefont {Fuentealba}}, \bibinfo {author} {\bibfnamefont {L.}~\bibnamefont {Gagliardi}}, \bibinfo {author} {\bibfnamefont {G.}~\bibnamefont {Galli}}, \bibinfo {author} {\bibfnamefont {J.}~\bibnamefont {Gao}}, \bibinfo {author} {\bibfnamefont {P.}~\bibnamefont {Geerlings}}, \bibinfo {author} {\bibfnamefont {N.}~\bibnamefont {Gidopoulos}}, \bibinfo {author} {\bibfnamefont {P.~M.~W.}\ \bibnamefont {Gill}}, \bibinfo {author} {\bibfnamefont {P.}~\bibnamefont {Gori-Giorgi}}, \bibinfo {author} {\bibfnamefont {A.}~\bibnamefont {G{\"o}rling}}, \bibinfo {author} {\bibfnamefont {T.}~\bibnamefont {Gould}}, \bibinfo {author}
  {\bibfnamefont {S.}~\bibnamefont {Grimme}}, \bibinfo {author} {\bibfnamefont {O.}~\bibnamefont {Gritsenko}}, \bibinfo {author} {\bibfnamefont {H.~J.~A.}\ \bibnamefont {Jensen}}, \bibinfo {author} {\bibfnamefont {E.~R.}\ \bibnamefont {Johnson}}, \bibinfo {author} {\bibfnamefont {R.~O.}\ \bibnamefont {Jones}}, \bibinfo {author} {\bibfnamefont {M.}~\bibnamefont {Kaupp}}, \bibinfo {author} {\bibfnamefont {A.~M.}\ \bibnamefont {K{\"o}ster}}, \bibinfo {author} {\bibfnamefont {L.}~\bibnamefont {Kronik}}, \bibinfo {author} {\bibfnamefont {A.~I.}\ \bibnamefont {Krylov}}, \bibinfo {author} {\bibfnamefont {S.}~\bibnamefont {Kvaal}}, \bibinfo {author} {\bibfnamefont {A.}~\bibnamefont {Laestadius}}, \bibinfo {author} {\bibfnamefont {M.}~\bibnamefont {Levy}}, \bibinfo {author} {\bibfnamefont {M.}~\bibnamefont {Lewin}}, \bibinfo {author} {\bibfnamefont {S.}~\bibnamefont {Liu}}, \bibinfo {author} {\bibfnamefont {P.-F.}\ \bibnamefont {Loos}}, \bibinfo {author} {\bibfnamefont {N.~T.}\ \bibnamefont {Maitra}}, \bibinfo {author} {\bibfnamefont {F.}~\bibnamefont {Neese}}, \bibinfo {author} {\bibfnamefont {J.~P.}\ \bibnamefont {Perdew}}, \bibinfo {author} {\bibfnamefont {K.}~\bibnamefont {Pernal}}, \bibinfo {author} {\bibfnamefont {P.}~\bibnamefont {Pernot}}, \bibinfo {author} {\bibfnamefont {P.}~\bibnamefont {Piecuch}}, \bibinfo {author} {\bibfnamefont {E.}~\bibnamefont {Rebolini}}, \bibinfo {author} {\bibfnamefont {L.}~\bibnamefont {Reining}}, \bibinfo {author} {\bibfnamefont {P.}~\bibnamefont {Romaniello}}, \bibinfo {author} {\bibfnamefont {A.}~\bibnamefont {Ruzsinszky}}, \bibinfo {author} {\bibfnamefont {D.~R.}\ \bibnamefont {Salahub}}, \bibinfo {author} {\bibfnamefont {M.}~\bibnamefont {Scheffler}}, \bibinfo {author} {\bibfnamefont {P.}~\bibnamefont {Schwerdtfeger}}, \bibinfo {author} {\bibfnamefont {V.~N.}\ \bibnamefont {Staroverov}}, \bibinfo {author} {\bibfnamefont {J.}~\bibnamefont {Sun}}, \bibinfo {author} {\bibfnamefont {E.}~\bibnamefont {Tellgren}}, \bibinfo {author} {\bibfnamefont {D.~J.}\ \bibnamefont {Tozer}}, \bibinfo
  {author} {\bibfnamefont {S.~B.}\ \bibnamefont {Trickey}}, \bibinfo {author} {\bibfnamefont {C.~A.}\ \bibnamefont {Ullrich}}, \bibinfo {author} {\bibfnamefont {A.}~\bibnamefont {Vela}}, \bibinfo {author} {\bibfnamefont {G.}~\bibnamefont {Vignale}}, \bibinfo {author} {\bibfnamefont {T.~A.}\ \bibnamefont {Wesolowski}}, \bibinfo {author} {\bibfnamefont {X.}~\bibnamefont {Xu}}, \ and\ \bibinfo {author} {\bibfnamefont {W.}~\bibnamefont {Yang}},\ }\bibfield  {title} {\enquote {\bibinfo {title} {Dft exchange: sharing perspectives on the workhorse of quantum chemistry and materials science},}\ }\href {\doibase 10.1039/D2CP02827A} {\bibfield  {journal} {\bibinfo  {journal} {Phys. Chem. Chem. Phys.}\ }\textbf {\bibinfo {volume} {24}},\ \bibinfo {pages} {28700--28781} (\bibinfo {year} {2022})}\BibitemShut {NoStop}%
\bibitem [{\citenamefont {Oudot}\ and\ \citenamefont {Doblhoff-Dier}(2024)}]{Oudot2024}%
  \BibitemOpen
  \bibfield  {author} {\bibinfo {author} {\bibfnamefont {B.}~\bibnamefont {Oudot}}\ and\ \bibinfo {author} {\bibfnamefont {K.}~\bibnamefont {Doblhoff-Dier}},\ }\bibfield  {title} {\enquote {\bibinfo {title} {Reaction barriers at metal surfaces computed using the random phase approximation: Can we beat dft in the generalized gradient approximation?}}\ }\href@noop {} {\bibfield  {journal} {\bibinfo  {journal} {J. Chem. Phys.}\ ,\ \bibinfo {pages} {054708}} (\bibinfo {year} {2024})}\BibitemShut {NoStop}%
\bibitem [{\citenamefont {Radoń}(2024)}]{Radon2024}%
  \BibitemOpen
  \bibfield  {author} {\bibinfo {author} {\bibfnamefont {M.}~\bibnamefont {Radoń}},\ }\bibfield  {title} {\enquote {\bibinfo {title} {Predicting spin states of iron porphyrins with dft methods including crystal packing effects and thermodynamic corrections},}\ }\href {\doibase 10.1039/D4CP01327A} {\bibfield  {journal} {\bibinfo  {journal} {Phys. Chem. Chem. Phys.}\ }\textbf {\bibinfo {volume} {26}},\ \bibinfo {pages} {18182--18195} (\bibinfo {year} {2024})}\BibitemShut {NoStop}%
\bibitem [{\citenamefont {Feynman}(1982)}]{Feynman1981}%
  \BibitemOpen
  \bibfield  {author} {\bibinfo {author} {\bibfnamefont {R.~P.}\ \bibnamefont {Feynman}},\ }\bibfield  {title} {\enquote {\bibinfo {title} {Simulating physics with computers},}\ }\href@noop {} {\bibfield  {journal} {\bibinfo  {journal} {International Journal of Theoretical Physics}\ }\textbf {\bibinfo {volume} {21}} (\bibinfo {year} {1982})}\BibitemShut {NoStop}%
\bibitem [{\citenamefont {Abrams}\ and\ \citenamefont {Lloyd}(1999)}]{Abrams1999}%
  \BibitemOpen
  \bibfield  {author} {\bibinfo {author} {\bibfnamefont {D.~S.}\ \bibnamefont {Abrams}}\ and\ \bibinfo {author} {\bibfnamefont {S.}~\bibnamefont {Lloyd}},\ }\bibfield  {title} {\enquote {\bibinfo {title} {Quantum algorithm providing exponential speed increase for finding eigenvalues and eigenvectors},}\ }\href {\doibase 10.1103/PhysRevLett.83.5162} {\bibfield  {journal} {\bibinfo  {journal} {Phys. Rev. Lett.}\ }\textbf {\bibinfo {volume} {83}},\ \bibinfo {pages} {5162--5165} (\bibinfo {year} {1999})}\BibitemShut {NoStop}%
\bibitem [{\citenamefont {Blunt}, \citenamefont {Geh\'er},\ and\ \citenamefont {Moylett}(2024)}]{Blunt2024}%
  \BibitemOpen
  \bibfield  {author} {\bibinfo {author} {\bibfnamefont {N.~S.}\ \bibnamefont {Blunt}}, \bibinfo {author} {\bibfnamefont {G.~P.}\ \bibnamefont {Geh\'er}}, \ and\ \bibinfo {author} {\bibfnamefont {A.~E.}\ \bibnamefont {Moylett}},\ }\bibfield  {title} {\enquote {\bibinfo {title} {Compilation of a simple chemistry application to quantum error correction primitives},}\ }\href {\doibase 10.1103/PhysRevResearch.6.013325} {\bibfield  {journal} {\bibinfo  {journal} {Phys. Rev. Res.}\ }\textbf {\bibinfo {volume} {6}},\ \bibinfo {pages} {013325} (\bibinfo {year} {2024})}\BibitemShut {NoStop}%
\bibitem [{\citenamefont {van Dam}\ \emph {et~al.}(2024)\citenamefont {van Dam}, \citenamefont {Liu}, \citenamefont {Low}, \citenamefont {Paetznick}, \citenamefont {Paz}, \citenamefont {Silva}, \citenamefont {Sundaram}, \citenamefont {Svore},\ and\ \citenamefont {Troyer}}]{vandam2024}%
  \BibitemOpen
  \bibfield  {author} {\bibinfo {author} {\bibfnamefont {W.}~\bibnamefont {van Dam}}, \bibinfo {author} {\bibfnamefont {H.}~\bibnamefont {Liu}}, \bibinfo {author} {\bibfnamefont {G.~H.}\ \bibnamefont {Low}}, \bibinfo {author} {\bibfnamefont {A.}~\bibnamefont {Paetznick}}, \bibinfo {author} {\bibfnamefont {A.}~\bibnamefont {Paz}}, \bibinfo {author} {\bibfnamefont {M.}~\bibnamefont {Silva}}, \bibinfo {author} {\bibfnamefont {A.}~\bibnamefont {Sundaram}}, \bibinfo {author} {\bibfnamefont {K.}~\bibnamefont {Svore}}, \ and\ \bibinfo {author} {\bibfnamefont {M.}~\bibnamefont {Troyer}},\ }\href {https://arxiv.org/abs/2409.05835} {\enquote {\bibinfo {title} {End-to-end quantum simulation of a chemical system},}\ } (\bibinfo {year} {2024}),\ \Eprint {http://arxiv.org/abs/2409.05835} {arXiv:2409.05835 [quant-ph]} \BibitemShut {NoStop}%
\bibitem [{\citenamefont {Dobrautz}\ \emph {et~al.}(2024)\citenamefont {Dobrautz}, \citenamefont {Sokolov}, \citenamefont {Liao}, \citenamefont {Ríos}, \citenamefont {Rahm}, \citenamefont {Alavi},\ and\ \citenamefont {Tavernelli}}]{Dobrautz2024}%
  \BibitemOpen
  \bibfield  {author} {\bibinfo {author} {\bibfnamefont {W.}~\bibnamefont {Dobrautz}}, \bibinfo {author} {\bibfnamefont {I.~O.}\ \bibnamefont {Sokolov}}, \bibinfo {author} {\bibfnamefont {K.}~\bibnamefont {Liao}}, \bibinfo {author} {\bibfnamefont {P.~L.}\ \bibnamefont {Ríos}}, \bibinfo {author} {\bibfnamefont {M.}~\bibnamefont {Rahm}}, \bibinfo {author} {\bibfnamefont {A.}~\bibnamefont {Alavi}}, \ and\ \bibinfo {author} {\bibfnamefont {I.}~\bibnamefont {Tavernelli}},\ }\bibfield  {title} {\enquote {\bibinfo {title} {Toward real chemical accuracy on current quantum hardware through the transcorrelated method},}\ }\href {\doibase 10.1021/acs.jctc.4c00070} {\bibfield  {journal} {\bibinfo  {journal} {Journal of Chemical Theory and Computation}\ }\textbf {\bibinfo {volume} {20}},\ \bibinfo {pages} {4146--4160} (\bibinfo {year} {2024})}\BibitemShut {NoStop}%
\bibitem [{\citenamefont {Kandula}\ \emph {et~al.}(2010)\citenamefont {Kandula}, \citenamefont {Gohle}, \citenamefont {Pinkert}, \citenamefont {Ubachs},\ and\ \citenamefont {Eikema}}]{Kandula2010}%
  \BibitemOpen
  \bibfield  {author} {\bibinfo {author} {\bibfnamefont {D.~Z.}\ \bibnamefont {Kandula}}, \bibinfo {author} {\bibfnamefont {C.}~\bibnamefont {Gohle}}, \bibinfo {author} {\bibfnamefont {T.~J.}\ \bibnamefont {Pinkert}}, \bibinfo {author} {\bibfnamefont {W.}~\bibnamefont {Ubachs}}, \ and\ \bibinfo {author} {\bibfnamefont {K.~S.~E.}\ \bibnamefont {Eikema}},\ }\bibfield  {title} {\enquote {\bibinfo {title} {Extreme ultraviolet frequency comb metrology},}\ }\href {\doibase 10.1103/PhysRevLett.105.063001} {\bibfield  {journal} {\bibinfo  {journal} {Phys. Rev. Lett.}\ }\textbf {\bibinfo {volume} {105}},\ \bibinfo {pages} {063001} (\bibinfo {year} {2010})}\BibitemShut {NoStop}%
\bibitem [{\citenamefont {Cao}\ \emph {et~al.}(2019)\citenamefont {Cao}, \citenamefont {Romero}, \citenamefont {Olson}, \citenamefont {Degroote}, \citenamefont {Johnson}, \citenamefont {Kieferová}, \citenamefont {Kivlichan}, \citenamefont {Menke}, \citenamefont {Peropadre}, \citenamefont {Sawaya}, \citenamefont {Sim}, \citenamefont {Veis},\ and\ \citenamefont {Aspuru-Guzik}}]{Cao2019}%
  \BibitemOpen
  \bibfield  {author} {\bibinfo {author} {\bibfnamefont {Y.}~\bibnamefont {Cao}}, \bibinfo {author} {\bibfnamefont {J.}~\bibnamefont {Romero}}, \bibinfo {author} {\bibfnamefont {J.~P.}\ \bibnamefont {Olson}}, \bibinfo {author} {\bibfnamefont {M.}~\bibnamefont {Degroote}}, \bibinfo {author} {\bibfnamefont {P.~D.}\ \bibnamefont {Johnson}}, \bibinfo {author} {\bibfnamefont {M.}~\bibnamefont {Kieferová}}, \bibinfo {author} {\bibfnamefont {I.~D.}\ \bibnamefont {Kivlichan}}, \bibinfo {author} {\bibfnamefont {T.}~\bibnamefont {Menke}}, \bibinfo {author} {\bibfnamefont {B.}~\bibnamefont {Peropadre}}, \bibinfo {author} {\bibfnamefont {N.~P.~D.}\ \bibnamefont {Sawaya}}, \bibinfo {author} {\bibfnamefont {S.}~\bibnamefont {Sim}}, \bibinfo {author} {\bibfnamefont {L.}~\bibnamefont {Veis}}, \ and\ \bibinfo {author} {\bibfnamefont {A.}~\bibnamefont {Aspuru-Guzik}},\ }\bibfield  {title} {\enquote {\bibinfo {title} {Quantum chemistry in the age of quantum computing},}\ }\href@noop {} {\bibfield  {journal} {\bibinfo  {journal} {Chem. Rev.}\ }\textbf {\bibinfo {volume} {119}},\ \bibinfo {pages} {10856--19015} (\bibinfo {year} {2019})}\BibitemShut {NoStop}%
\bibitem [{\citenamefont {Setia}\ \emph {et~al.}(2020)\citenamefont {Setia}, \citenamefont {Chen}, \citenamefont {Rice}, \citenamefont {Mezzacapo}, \citenamefont {Pistoia},\ and\ \citenamefont {Whitfield}}]{Setia2020}%
  \BibitemOpen
  \bibfield  {author} {\bibinfo {author} {\bibfnamefont {K.}~\bibnamefont {Setia}}, \bibinfo {author} {\bibfnamefont {R.}~\bibnamefont {Chen}}, \bibinfo {author} {\bibfnamefont {J.~E.}\ \bibnamefont {Rice}}, \bibinfo {author} {\bibfnamefont {A.}~\bibnamefont {Mezzacapo}}, \bibinfo {author} {\bibfnamefont {M.}~\bibnamefont {Pistoia}}, \ and\ \bibinfo {author} {\bibfnamefont {J.~D.}\ \bibnamefont {Whitfield}},\ }\bibfield  {title} {\enquote {\bibinfo {title} {Reducing qubit requirements for quantum simulations using molecular point group symmetries},}\ }\href@noop {} {\bibfield  {journal} {\bibinfo  {journal} {Journal of Chemical Theory and Computation}\ ,\ \bibinfo {pages} {6091--6097}} (\bibinfo {year} {2020})}\BibitemShut {NoStop}%
\bibitem [{\citenamefont {Babbush}\ \emph {et~al.}(2018)\citenamefont {Babbush}, \citenamefont {Berry}, \citenamefont {Sanders}, \citenamefont {Kivlichan}, \citenamefont {Scherer}, \citenamefont {Wei}, \citenamefont {Love},\ and\ \citenamefont {Aspuru-Guzik}}]{Babbush2018}%
  \BibitemOpen
  \bibfield  {author} {\bibinfo {author} {\bibfnamefont {R.}~\bibnamefont {Babbush}}, \bibinfo {author} {\bibfnamefont {D.~W.}\ \bibnamefont {Berry}}, \bibinfo {author} {\bibfnamefont {Y.~R.}\ \bibnamefont {Sanders}}, \bibinfo {author} {\bibfnamefont {I.~D.}\ \bibnamefont {Kivlichan}}, \bibinfo {author} {\bibfnamefont {A.}~\bibnamefont {Scherer}}, \bibinfo {author} {\bibfnamefont {A.~Y.}\ \bibnamefont {Wei}}, \bibinfo {author} {\bibfnamefont {P.~J.}\ \bibnamefont {Love}}, \ and\ \bibinfo {author} {\bibfnamefont {A.}~\bibnamefont {Aspuru-Guzik}},\ }\bibfield  {title} {\enquote {\bibinfo {title} {Exponentially more precise quantum simulation of fermions in the configuration interaction representation},}\ }\href@noop {} {\bibfield  {journal} {\bibinfo  {journal} {Quantum Sci. Technol.}\ ,\ \bibinfo {pages} {015006}} (\bibinfo {year} {2018})}\BibitemShut {NoStop}%
\bibitem [{\citenamefont {Bonet-Monroig}\ \emph {et~al.}(2018)\citenamefont {Bonet-Monroig}, \citenamefont {Sagastizabal}, \citenamefont {Singh},\ and\ \citenamefont {O'Brien}}]{Bonet-Monroig2018}%
  \BibitemOpen
  \bibfield  {author} {\bibinfo {author} {\bibfnamefont {X.}~\bibnamefont {Bonet-Monroig}}, \bibinfo {author} {\bibfnamefont {R.}~\bibnamefont {Sagastizabal}}, \bibinfo {author} {\bibfnamefont {M.}~\bibnamefont {Singh}}, \ and\ \bibinfo {author} {\bibfnamefont {T.~E.}\ \bibnamefont {O'Brien}},\ }\bibfield  {title} {\enquote {\bibinfo {title} {Low-cost error mitigation by symmetry verification},}\ }\href {\doibase 10.1103/PhysRevA.98.062339} {\bibfield  {journal} {\bibinfo  {journal} {Phys. Rev. A}\ }\textbf {\bibinfo {volume} {98}},\ \bibinfo {pages} {062339} (\bibinfo {year} {2018})}\BibitemShut {NoStop}%
\bibitem [{\citenamefont {Szab{\'o}}\ and\ \citenamefont {Ostlund}(1996)}]{Szabo}%
  \BibitemOpen
  \bibfield  {author} {\bibinfo {author} {\bibfnamefont {A.}~\bibnamefont {Szab{\'o}}}\ and\ \bibinfo {author} {\bibfnamefont {N.~S.}\ \bibnamefont {Ostlund}},\ }\href {http://lib.ugent.be/catalog/rug01:000906565} {\emph {\bibinfo {title} {Modern quantum chemistry : introduction to advanced electronic structure theory}}}\ (\bibinfo  {publisher} {Mineola (N.Y.) : Dover publications},\ \bibinfo {year} {1996})\BibitemShut {NoStop}%
\bibitem [{\citenamefont {Volkmann}\ \emph {et~al.}(2024)\citenamefont {Volkmann}, \citenamefont {Sathyanarayanan}, \citenamefont {Saenz}, \citenamefont {Jansen},\ and\ \citenamefont {K\"uhn}}]{Volkmann2024}%
  \BibitemOpen
  \bibfield  {author} {\bibinfo {author} {\bibfnamefont {H.}~\bibnamefont {Volkmann}}, \bibinfo {author} {\bibfnamefont {R.}~\bibnamefont {Sathyanarayanan}}, \bibinfo {author} {\bibfnamefont {A.}~\bibnamefont {Saenz}}, \bibinfo {author} {\bibfnamefont {K.}~\bibnamefont {Jansen}}, \ and\ \bibinfo {author} {\bibfnamefont {S.}~\bibnamefont {K\"uhn}},\ }\bibfield  {title} {\enquote {\bibinfo {title} {Chemically accurate potential curves for {H}$_2$ molecules using explicitly correlated qubit-adapt},}\ }\href@noop {} {\bibfield  {journal} {\bibinfo  {journal} {J. Chem. Theory Comput.}\ }\textbf {\bibinfo {volume} {20}},\ \bibinfo {pages} {1244--1251} (\bibinfo {year} {2024})}\BibitemShut {NoStop}%
\bibitem [{\citenamefont {Elfving}\ \emph {et~al.}(2021)\citenamefont {Elfving}, \citenamefont {Millaruelo}, \citenamefont {G\'amez},\ and\ \citenamefont {Gogolin}}]{Elfving2021}%
  \BibitemOpen
  \bibfield  {author} {\bibinfo {author} {\bibfnamefont {V.~E.}\ \bibnamefont {Elfving}}, \bibinfo {author} {\bibfnamefont {M.}~\bibnamefont {Millaruelo}}, \bibinfo {author} {\bibfnamefont {J.~A.}\ \bibnamefont {G\'amez}}, \ and\ \bibinfo {author} {\bibfnamefont {C.}~\bibnamefont {Gogolin}},\ }\bibfield  {title} {\enquote {\bibinfo {title} {Simulating quantum chemistry in the seniority-zero space on qubit-based quantum computers},}\ }\href {\doibase 10.1103/PhysRevA.103.032605} {\bibfield  {journal} {\bibinfo  {journal} {Phys. Rev. A}\ }\textbf {\bibinfo {volume} {103}},\ \bibinfo {pages} {032605} (\bibinfo {year} {2021})}\BibitemShut {NoStop}%
\bibitem [{\citenamefont {Zhao}\ \emph {et~al.}(2023)\citenamefont {Zhao}, \citenamefont {Goings}, \citenamefont {Shin}, \citenamefont {Kyoung}, \citenamefont {Fuks}, \citenamefont {Kevin~Rhee}, \citenamefont {Rhee}, \citenamefont {Wright}, \citenamefont {Nguyen}, \citenamefont {Kim},\ and\ \citenamefont {Johri}}]{Zhao2023}%
  \BibitemOpen
  \bibfield  {author} {\bibinfo {author} {\bibfnamefont {L.}~\bibnamefont {Zhao}}, \bibinfo {author} {\bibfnamefont {J.}~\bibnamefont {Goings}}, \bibinfo {author} {\bibfnamefont {K.}~\bibnamefont {Shin}}, \bibinfo {author} {\bibfnamefont {W.}~\bibnamefont {Kyoung}}, \bibinfo {author} {\bibfnamefont {J.~I.}\ \bibnamefont {Fuks}}, \bibinfo {author} {\bibfnamefont {J.-K.}\ \bibnamefont {Kevin~Rhee}}, \bibinfo {author} {\bibfnamefont {Y.~M.}\ \bibnamefont {Rhee}}, \bibinfo {author} {\bibfnamefont {K.}~\bibnamefont {Wright}}, \bibinfo {author} {\bibfnamefont {J.}~\bibnamefont {Nguyen}}, \bibinfo {author} {\bibfnamefont {J.}~\bibnamefont {Kim}}, \ and\ \bibinfo {author} {\bibfnamefont {S.}~\bibnamefont {Johri}},\ }\bibfield  {title} {\enquote {\bibinfo {title} {Orbital-optimized pair-correlated electron simulations on trapped-ion quantum computers},}\ }\href {\doibase 10.1038/s41534-023-00730-8} {\bibfield  {journal} {\bibinfo  {journal} {npj Quantum Information}\ }\textbf {\bibinfo {volume} {9}},\ \bibinfo {pages} {60} (\bibinfo {year} {2023})}\BibitemShut {NoStop}%
\bibitem [{\citenamefont {Sawaya}\ \emph {et~al.}(2020)\citenamefont {Sawaya}, \citenamefont {Menke}, \citenamefont {Kyaw}, \citenamefont {Johri}, \citenamefont {Aspuru-Guzik},\ and\ \citenamefont {Guerreschi}}]{Sawaya2020}%
  \BibitemOpen
  \bibfield  {author} {\bibinfo {author} {\bibfnamefont {N.~P.~D.}\ \bibnamefont {Sawaya}}, \bibinfo {author} {\bibfnamefont {T.}~\bibnamefont {Menke}}, \bibinfo {author} {\bibfnamefont {T.~H.}\ \bibnamefont {Kyaw}}, \bibinfo {author} {\bibfnamefont {S.}~\bibnamefont {Johri}}, \bibinfo {author} {\bibfnamefont {A.}~\bibnamefont {Aspuru-Guzik}}, \ and\ \bibinfo {author} {\bibfnamefont {G.~G.}\ \bibnamefont {Guerreschi}},\ }\bibfield  {title} {\enquote {\bibinfo {title} {Resource-efficient digital quantum simulation of d-level systems for photonic, vibrational, and spin-s hamiltonians},}\ }\href {\doibase 10.1038/s41534-020-0278-0} {\bibfield  {journal} {\bibinfo  {journal} {npj Quantum Information}\ }\textbf {\bibinfo {volume} {6}},\ \bibinfo {pages} {49} (\bibinfo {year} {2020})}\BibitemShut {NoStop}%
\bibitem [{\citenamefont {Peruzzo}\ \emph {et~al.}(2014)\citenamefont {Peruzzo}, \citenamefont {McClean}, \citenamefont {Shadbolt}, \citenamefont {Yung}, \citenamefont {Zhou}, \citenamefont {Love}, \citenamefont {Aspuru-Guzik},\ and\ \citenamefont {O'Brien}}]{Peruzzo2014}%
  \BibitemOpen
  \bibfield  {author} {\bibinfo {author} {\bibfnamefont {A.}~\bibnamefont {Peruzzo}}, \bibinfo {author} {\bibfnamefont {J.}~\bibnamefont {McClean}}, \bibinfo {author} {\bibfnamefont {P.}~\bibnamefont {Shadbolt}}, \bibinfo {author} {\bibfnamefont {M.-H.}\ \bibnamefont {Yung}}, \bibinfo {author} {\bibfnamefont {X.-Q.}\ \bibnamefont {Zhou}}, \bibinfo {author} {\bibfnamefont {P.~J.}\ \bibnamefont {Love}}, \bibinfo {author} {\bibfnamefont {A.}~\bibnamefont {Aspuru-Guzik}}, \ and\ \bibinfo {author} {\bibfnamefont {J.~L.}\ \bibnamefont {O'Brien}},\ }\bibfield  {title} {\enquote {\bibinfo {title} {A variational eigenvalue solver on a photonic quantum processor},}\ }\href {\doibase 10.1038/ncomms5213} {\bibfield  {journal} {\bibinfo  {journal} {Nature Communications}\ }\textbf {\bibinfo {volume} {5}},\ \bibinfo {pages} {4213} (\bibinfo {year} {2014})}\BibitemShut {NoStop}%
\bibitem [{\citenamefont {Ostaszewski}, \citenamefont {Grant},\ and\ \citenamefont {Benedetti}(2021)}]{Ostaszewski2021}%
  \BibitemOpen
  \bibfield  {author} {\bibinfo {author} {\bibfnamefont {M.}~\bibnamefont {Ostaszewski}}, \bibinfo {author} {\bibfnamefont {E.}~\bibnamefont {Grant}}, \ and\ \bibinfo {author} {\bibfnamefont {M.}~\bibnamefont {Benedetti}},\ }\bibfield  {title} {\enquote {\bibinfo {title} {Structure optimization for parameterized quantum circuits},}\ }\href@noop {} {\bibfield  {journal} {\bibinfo  {journal} {Quantum}\ ,\ \bibinfo {pages} {391}} (\bibinfo {year} {2021})}\BibitemShut {NoStop}%
\bibitem [{\citenamefont {Vallury}\ \emph {et~al.}(2020)\citenamefont {Vallury}, \citenamefont {Jones}, \citenamefont {Hill},\ and\ \citenamefont {Hollenberg}}]{Vallury2020}%
  \BibitemOpen
  \bibfield  {author} {\bibinfo {author} {\bibfnamefont {H.~J.}\ \bibnamefont {Vallury}}, \bibinfo {author} {\bibfnamefont {M.~A.}\ \bibnamefont {Jones}}, \bibinfo {author} {\bibfnamefont {C.~D.}\ \bibnamefont {Hill}}, \ and\ \bibinfo {author} {\bibfnamefont {L.~C.~L.}\ \bibnamefont {Hollenberg}},\ }\bibfield  {title} {\enquote {\bibinfo {title} {Quantum computed moments correction to variational estimates},}\ }\href@noop {} {\bibfield  {journal} {\bibinfo  {journal} {Quantum}\ ,\ \bibinfo {pages} {373}} (\bibinfo {year} {2020})}\BibitemShut {NoStop}%
\bibitem [{\citenamefont {Hollenberg}(1993)}]{Hollenberg1993}%
  \BibitemOpen
  \bibfield  {author} {\bibinfo {author} {\bibfnamefont {L.~C.~L.}\ \bibnamefont {Hollenberg}},\ }\bibfield  {title} {\enquote {\bibinfo {title} {Plaquette expansion in lattice hamiltonian models},}\ }\href {\doibase 10.1103/PhysRevD.47.1640} {\bibfield  {journal} {\bibinfo  {journal} {Phys. Rev. D}\ }\textbf {\bibinfo {volume} {47}},\ \bibinfo {pages} {1640--1644} (\bibinfo {year} {1993})}\BibitemShut {NoStop}%
\bibitem [{\citenamefont {Witte}\ and\ \citenamefont {Hollenberg}(1997)}]{Witte1997}%
  \BibitemOpen
  \bibfield  {author} {\bibinfo {author} {\bibfnamefont {N.~S.}\ \bibnamefont {Witte}}\ and\ \bibinfo {author} {\bibfnamefont {L.~C.~L.}\ \bibnamefont {Hollenberg}},\ }\bibfield  {title} {\enquote {\bibinfo {title} {Accurate calculation of ground-state energies in an analytic lanczos expansion},}\ }\href {\doibase 10.1088/0953-8984/9/9/016} {\bibfield  {journal} {\bibinfo  {journal} {Journal of Physics: Condensed Matter}\ }\textbf {\bibinfo {volume} {9}},\ \bibinfo {pages} {2031} (\bibinfo {year} {1997})}\BibitemShut {NoStop}%
\bibitem [{\citenamefont {Hollenberg}\ and\ \citenamefont {Witte}(1994)}]{Hollenberg1994}%
  \BibitemOpen
  \bibfield  {author} {\bibinfo {author} {\bibfnamefont {L.~C.~L.}\ \bibnamefont {Hollenberg}}\ and\ \bibinfo {author} {\bibfnamefont {N.~S.}\ \bibnamefont {Witte}},\ }\bibfield  {title} {\enquote {\bibinfo {title} {General nonperturbative estimate of the energy density of lattice hamiltonians},}\ }\href {\doibase 10.1103/PhysRevD.50.3382} {\bibfield  {journal} {\bibinfo  {journal} {Phys. Rev. D}\ }\textbf {\bibinfo {volume} {50}},\ \bibinfo {pages} {3382--3386} (\bibinfo {year} {1994})}\BibitemShut {NoStop}%
\bibitem [{\citenamefont {Jones}\ \emph {et~al.}(2022)\citenamefont {Jones}, \citenamefont {Vallury}, \citenamefont {Hill},\ and\ \citenamefont {Hollenberg}}]{Jones2022}%
  \BibitemOpen
  \bibfield  {author} {\bibinfo {author} {\bibfnamefont {M.~A.}\ \bibnamefont {Jones}}, \bibinfo {author} {\bibfnamefont {H.~J.}\ \bibnamefont {Vallury}}, \bibinfo {author} {\bibfnamefont {C.~D.}\ \bibnamefont {Hill}}, \ and\ \bibinfo {author} {\bibfnamefont {L.~C.~L.}\ \bibnamefont {Hollenberg}},\ }\bibfield  {title} {\enquote {\bibinfo {title} {Chemistry beyond the hartree--fock energy via quantum computed moments},}\ }\href {\doibase 10.1038/s41598-022-12324-z} {\bibfield  {journal} {\bibinfo  {journal} {Scientific Reports}\ }\textbf {\bibinfo {volume} {12}},\ \bibinfo {pages} {8985} (\bibinfo {year} {2022})}\BibitemShut {NoStop}%
\bibitem [{\citenamefont {Jones}, \citenamefont {Vallury},\ and\ \citenamefont {Hollenberg}(2024)}]{Jones2024}%
  \BibitemOpen
  \bibfield  {author} {\bibinfo {author} {\bibfnamefont {M.~A.}\ \bibnamefont {Jones}}, \bibinfo {author} {\bibfnamefont {H.~J.}\ \bibnamefont {Vallury}}, \ and\ \bibinfo {author} {\bibfnamefont {L.~C.}\ \bibnamefont {Hollenberg}},\ }\bibfield  {title} {\enquote {\bibinfo {title} {Ground-state-energy calculation for the water molecule on a superconducting quantum processor},}\ }\href {\doibase 10.1103/PhysRevApplied.21.064017} {\bibfield  {journal} {\bibinfo  {journal} {Phys. Rev. Appl.}\ }\textbf {\bibinfo {volume} {21}},\ \bibinfo {pages} {064017} (\bibinfo {year} {2024})}\BibitemShut {NoStop}%
\bibitem [{\citenamefont {Greene-Diniz}\ \emph {et~al.}(2024)\citenamefont {Greene-Diniz}, \citenamefont {Manrique}, \citenamefont {Yamamoto}, \citenamefont {Plekhanov}, \citenamefont {Fitzpatrick}, \citenamefont {Krompiec}, \citenamefont {Sakuma},\ and\ \citenamefont {Ramo}}]{GreeneDiniz2024}%
  \BibitemOpen
  \bibfield  {author} {\bibinfo {author} {\bibfnamefont {G.}~\bibnamefont {Greene-Diniz}}, \bibinfo {author} {\bibfnamefont {D.~Z.}\ \bibnamefont {Manrique}}, \bibinfo {author} {\bibfnamefont {K.}~\bibnamefont {Yamamoto}}, \bibinfo {author} {\bibfnamefont {E.}~\bibnamefont {Plekhanov}}, \bibinfo {author} {\bibfnamefont {N.}~\bibnamefont {Fitzpatrick}}, \bibinfo {author} {\bibfnamefont {M.}~\bibnamefont {Krompiec}}, \bibinfo {author} {\bibfnamefont {R.}~\bibnamefont {Sakuma}}, \ and\ \bibinfo {author} {\bibfnamefont {D.~M.}\ \bibnamefont {Ramo}},\ }\bibfield  {title} {\enquote {\bibinfo {title} {Quantum {C}omputed {G}reen's {F}unctions using a {C}umulant {E}xpansion of the {L}anczos {M}ethod},}\ }\href {\doibase 10.22331/q-2024-06-20-1383} {\bibfield  {journal} {\bibinfo  {journal} {{Quantum}}\ }\textbf {\bibinfo {volume} {8}},\ \bibinfo {pages} {1383} (\bibinfo {year} {2024})}\BibitemShut {NoStop}%
\bibitem [{\citenamefont {Toloui}\ and\ \citenamefont {Love}(2013)}]{Toloui2013}%
  \BibitemOpen
  \bibfield  {author} {\bibinfo {author} {\bibfnamefont {B.}~\bibnamefont {Toloui}}\ and\ \bibinfo {author} {\bibfnamefont {P.~J.}\ \bibnamefont {Love}},\ }\href {https://arxiv.org/abs/1312.2579} {\enquote {\bibinfo {title} {Quantum algorithms for quantum chemistry based on the sparsity of the {CI}-matrix},}\ } (\bibinfo {year} {2013}),\ \Eprint {http://arxiv.org/abs/1312.2579} {arXiv:1312.2579 [quant-ph]} \BibitemShut {NoStop}%
\bibitem [{\citenamefont {Kirby}\ and\ \citenamefont {Love}(2021)}]{Kirby2021}%
  \BibitemOpen
  \bibfield  {author} {\bibinfo {author} {\bibfnamefont {W.~M.}\ \bibnamefont {Kirby}}\ and\ \bibinfo {author} {\bibfnamefont {P.~J.}\ \bibnamefont {Love}},\ }\bibfield  {title} {\enquote {\bibinfo {title} {Variational quantum eigensolvers for sparse hamiltonians},}\ }\href {\doibase 10.1103/PhysRevLett.127.110503} {\bibfield  {journal} {\bibinfo  {journal} {Phys. Rev. Lett.}\ }\textbf {\bibinfo {volume} {127}},\ \bibinfo {pages} {110503} (\bibinfo {year} {2021})}\BibitemShut {NoStop}%
\bibitem [{\citenamefont {Temme}, \citenamefont {Bravyi},\ and\ \citenamefont {Gambetta}(2017)}]{Temme2017}%
  \BibitemOpen
  \bibfield  {author} {\bibinfo {author} {\bibfnamefont {K.}~\bibnamefont {Temme}}, \bibinfo {author} {\bibfnamefont {S.}~\bibnamefont {Bravyi}}, \ and\ \bibinfo {author} {\bibfnamefont {J.~M.}\ \bibnamefont {Gambetta}},\ }\bibfield  {title} {\enquote {\bibinfo {title} {Error mitigation for short-depth quantum circuits},}\ }\href {\doibase 10.1103/PhysRevLett.119.180509} {\bibfield  {journal} {\bibinfo  {journal} {Phys. Rev. Lett.}\ }\textbf {\bibinfo {volume} {119}},\ \bibinfo {pages} {180509} (\bibinfo {year} {2017})}\BibitemShut {NoStop}%
\bibitem [{\citenamefont {Huggins}\ \emph {et~al.}(2021)\citenamefont {Huggins}, \citenamefont {McArdle}, \citenamefont {O'Brien}, \citenamefont {Lee}, \citenamefont {Rubin}, \citenamefont {Boixo}, \citenamefont {Whaley}, \citenamefont {Babbush},\ and\ \citenamefont {McClean}}]{Huggins2021}%
  \BibitemOpen
  \bibfield  {author} {\bibinfo {author} {\bibfnamefont {W.~J.}\ \bibnamefont {Huggins}}, \bibinfo {author} {\bibfnamefont {S.}~\bibnamefont {McArdle}}, \bibinfo {author} {\bibfnamefont {T.~E.}\ \bibnamefont {O'Brien}}, \bibinfo {author} {\bibfnamefont {J.}~\bibnamefont {Lee}}, \bibinfo {author} {\bibfnamefont {N.~C.}\ \bibnamefont {Rubin}}, \bibinfo {author} {\bibfnamefont {S.}~\bibnamefont {Boixo}}, \bibinfo {author} {\bibfnamefont {K.~B.}\ \bibnamefont {Whaley}}, \bibinfo {author} {\bibfnamefont {R.}~\bibnamefont {Babbush}}, \ and\ \bibinfo {author} {\bibfnamefont {J.~R.}\ \bibnamefont {McClean}},\ }\bibfield  {title} {\enquote {\bibinfo {title} {Virtual distillation for quantum error mitigation},}\ }\href {\doibase 10.1103/PhysRevX.11.041036} {\bibfield  {journal} {\bibinfo  {journal} {Phys. Rev. X}\ }\textbf {\bibinfo {volume} {11}},\ \bibinfo {pages} {041036} (\bibinfo {year} {2021})}\BibitemShut {NoStop}%
\bibitem [{\citenamefont {Smith}\ \emph {et~al.}(2018)\citenamefont {Smith}, \citenamefont {Burns}, \citenamefont {Sirianni}, \citenamefont {Nascimento}, \citenamefont {Kumar}, \citenamefont {James}, \citenamefont {Schriber}, \citenamefont {Zhang}, \citenamefont {Zhang}, \citenamefont {Abbott}, \citenamefont {Berquist}, \citenamefont {Lechner}, \citenamefont {Cunha}, \citenamefont {Heide}, \citenamefont {Waldrop}, \citenamefont {Takeshita}, \citenamefont {Alenaizan}, \citenamefont {Neuhauser}, \citenamefont {King}, \citenamefont {Simmonett}, \citenamefont {Turney}, \citenamefont {Schaefer}, \citenamefont {Evangelista}, \citenamefont {DePrince}, \citenamefont {Crawford}, \citenamefont {Patkowski},\ and\ \citenamefont {Sherrill}}]{Smith2018}%
  \BibitemOpen
  \bibfield  {author} {\bibinfo {author} {\bibfnamefont {D.~G.~A.}\ \bibnamefont {Smith}}, \bibinfo {author} {\bibfnamefont {L.~A.}\ \bibnamefont {Burns}}, \bibinfo {author} {\bibfnamefont {D.~A.}\ \bibnamefont {Sirianni}}, \bibinfo {author} {\bibfnamefont {D.~R.}\ \bibnamefont {Nascimento}}, \bibinfo {author} {\bibfnamefont {A.}~\bibnamefont {Kumar}}, \bibinfo {author} {\bibfnamefont {A.~M.}\ \bibnamefont {James}}, \bibinfo {author} {\bibfnamefont {J.~B.}\ \bibnamefont {Schriber}}, \bibinfo {author} {\bibfnamefont {T.}~\bibnamefont {Zhang}}, \bibinfo {author} {\bibfnamefont {B.}~\bibnamefont {Zhang}}, \bibinfo {author} {\bibfnamefont {A.~S.}\ \bibnamefont {Abbott}}, \bibinfo {author} {\bibfnamefont {E.~J.}\ \bibnamefont {Berquist}}, \bibinfo {author} {\bibfnamefont {M.~H.}\ \bibnamefont {Lechner}}, \bibinfo {author} {\bibfnamefont {L.~A.}\ \bibnamefont {Cunha}}, \bibinfo {author} {\bibfnamefont {A.~G.}\ \bibnamefont {Heide}}, \bibinfo {author} {\bibfnamefont {J.~M.}\ \bibnamefont {Waldrop}}, \bibinfo {author} {\bibfnamefont {T.~Y.}\ \bibnamefont {Takeshita}}, \bibinfo {author} {\bibfnamefont {A.}~\bibnamefont {Alenaizan}}, \bibinfo {author} {\bibfnamefont {D.}~\bibnamefont {Neuhauser}}, \bibinfo {author} {\bibfnamefont {R.~A.}\ \bibnamefont {King}}, \bibinfo {author} {\bibfnamefont {A.~C.}\ \bibnamefont {Simmonett}}, \bibinfo {author} {\bibfnamefont {J.~M.}\ \bibnamefont {Turney}}, \bibinfo {author} {\bibfnamefont {H.~F.}\ \bibnamefont {Schaefer}}, \bibinfo {author} {\bibfnamefont {F.~A.}\ \bibnamefont {Evangelista}}, \bibinfo {author} {\bibfnamefont {A.~E.}\ \bibnamefont {DePrince}}, \bibinfo {author} {\bibfnamefont {T.~D.}\ \bibnamefont {Crawford}}, \bibinfo {author} {\bibfnamefont {K.}~\bibnamefont {Patkowski}}, \ and\ \bibinfo {author} {\bibfnamefont {C.~D.}\ \bibnamefont {Sherrill}},\ }\bibfield  {title} {\enquote {\bibinfo {title} {Psi{4N}um{P}y : {An} {Interactive} {Quantum} {Chemistry} {Programming} {Environment} for {Reference} {Implementations} and {Rapid} {Development}},}\ }\href {\doibase
  10.1021/acs.jctc.8b00286} {\bibfield  {journal} {\bibinfo  {journal} {Journal of Chemical Theory and Computation}\ }\textbf {\bibinfo {volume} {14}},\ \bibinfo {pages} {3504--3511} (\bibinfo {year} {2018})}\BibitemShut {NoStop}%
\bibitem [{\citenamefont {Sun}\ \emph {et~al.}(2018)\citenamefont {Sun}, \citenamefont {Berkelbach}, \citenamefont {Blunt}, \citenamefont {Booth}, \citenamefont {Guo}, \citenamefont {Li}, \citenamefont {Liu}, \citenamefont {McClain}, \citenamefont {Sayfutyarova}, \citenamefont {Sharma}, \citenamefont {Wouters},\ and\ \citenamefont {Chan}}]{PySCF}%
  \BibitemOpen
  \bibfield  {author} {\bibinfo {author} {\bibfnamefont {Q.}~\bibnamefont {Sun}}, \bibinfo {author} {\bibfnamefont {T.~C.}\ \bibnamefont {Berkelbach}}, \bibinfo {author} {\bibfnamefont {N.~S.}\ \bibnamefont {Blunt}}, \bibinfo {author} {\bibfnamefont {G.~H.}\ \bibnamefont {Booth}}, \bibinfo {author} {\bibfnamefont {S.}~\bibnamefont {Guo}}, \bibinfo {author} {\bibfnamefont {Z.}~\bibnamefont {Li}}, \bibinfo {author} {\bibfnamefont {J.}~\bibnamefont {Liu}}, \bibinfo {author} {\bibfnamefont {J.~D.}\ \bibnamefont {McClain}}, \bibinfo {author} {\bibfnamefont {E.~R.}\ \bibnamefont {Sayfutyarova}}, \bibinfo {author} {\bibfnamefont {S.}~\bibnamefont {Sharma}}, \bibinfo {author} {\bibfnamefont {S.}~\bibnamefont {Wouters}}, \ and\ \bibinfo {author} {\bibfnamefont {G.~K.-L.}\ \bibnamefont {Chan}},\ }\bibfield  {title} {\enquote {\bibinfo {title} {Pyscf: the python-based simulations of chemistry framework},}\ }\href {\doibase https://doi.org/10.1002/wcms.1340} {\bibfield  {journal} {\bibinfo  {journal} {WIREs Computational Molecular Science}\ }\textbf {\bibinfo {volume} {8}},\ \bibinfo {pages} {e1340} (\bibinfo {year} {2018})}\BibitemShut {NoStop}%
\bibitem [{\citenamefont {Javadi-Abhari}\ \emph {et~al.}(2024)\citenamefont {Javadi-Abhari}, \citenamefont {Treinish}, \citenamefont {Krsulich}, \citenamefont {Wood}, \citenamefont {Lishman}, \citenamefont {Gacon}, \citenamefont {Martiel}, \citenamefont {Nation}, \citenamefont {Bishop}, \citenamefont {Cross}, \citenamefont {Johnson},\ and\ \citenamefont {Gambetta}}]{qiskit2024}%
  \BibitemOpen
  \bibfield  {author} {\bibinfo {author} {\bibfnamefont {A.}~\bibnamefont {Javadi-Abhari}}, \bibinfo {author} {\bibfnamefont {M.}~\bibnamefont {Treinish}}, \bibinfo {author} {\bibfnamefont {K.}~\bibnamefont {Krsulich}}, \bibinfo {author} {\bibfnamefont {C.~J.}\ \bibnamefont {Wood}}, \bibinfo {author} {\bibfnamefont {J.}~\bibnamefont {Lishman}}, \bibinfo {author} {\bibfnamefont {J.}~\bibnamefont {Gacon}}, \bibinfo {author} {\bibfnamefont {S.}~\bibnamefont {Martiel}}, \bibinfo {author} {\bibfnamefont {P.~D.}\ \bibnamefont {Nation}}, \bibinfo {author} {\bibfnamefont {L.~S.}\ \bibnamefont {Bishop}}, \bibinfo {author} {\bibfnamefont {A.~W.}\ \bibnamefont {Cross}}, \bibinfo {author} {\bibfnamefont {B.~R.}\ \bibnamefont {Johnson}}, \ and\ \bibinfo {author} {\bibfnamefont {J.~M.}\ \bibnamefont {Gambetta}},\ }\href {\doibase 10.48550/arXiv.2405.08810} {\enquote {\bibinfo {title} {Quantum computing with {Q}iskit},}\ } (\bibinfo {year} {2024}),\ \Eprint {http://arxiv.org/abs/2405.08810} {arXiv:2405.08810 [quant-ph]} \BibitemShut {NoStop}%
\bibitem [{\citenamefont {{Microsoft}}()}]{Microsoft_Azure_Quantum_Development}%
  \BibitemOpen
  \bibfield  {author} {\bibinfo {author} {\bibnamefont {{Microsoft}}},\ }\href {https://github.com/microsoft/qsharp} {\enquote {\bibinfo {title} {{Azure Quantum Development Kit}},}\ }\BibitemShut {NoStop}%
\bibitem [{\citenamefont {Svore}\ \emph {et~al.}(2018)\citenamefont {Svore}, \citenamefont {Geller}, \citenamefont {Troyer}, \citenamefont {Azariah}, \citenamefont {Granade}, \citenamefont {Heim}, \citenamefont {Kliuchnikov}, \citenamefont {Mykhailova}, \citenamefont {Paz},\ and\ \citenamefont {Roetteler}}]{Svore_2018}%
  \BibitemOpen
  \bibfield  {author} {\bibinfo {author} {\bibfnamefont {K.}~\bibnamefont {Svore}}, \bibinfo {author} {\bibfnamefont {A.}~\bibnamefont {Geller}}, \bibinfo {author} {\bibfnamefont {M.}~\bibnamefont {Troyer}}, \bibinfo {author} {\bibfnamefont {J.}~\bibnamefont {Azariah}}, \bibinfo {author} {\bibfnamefont {C.}~\bibnamefont {Granade}}, \bibinfo {author} {\bibfnamefont {B.}~\bibnamefont {Heim}}, \bibinfo {author} {\bibfnamefont {V.}~\bibnamefont {Kliuchnikov}}, \bibinfo {author} {\bibfnamefont {M.}~\bibnamefont {Mykhailova}}, \bibinfo {author} {\bibfnamefont {A.}~\bibnamefont {Paz}}, \ and\ \bibinfo {author} {\bibfnamefont {M.}~\bibnamefont {Roetteler}},\ }\bibfield  {title} {\enquote {\bibinfo {title} {{Q\#: Enabling Scalable Quantum Computing and Development with a High-level DSL}},}\ }in\ \href {\doibase 10.1145/3183895.3183901} {\emph {\bibinfo {booktitle} {Proceedings of the Real World Domain Specific Languages Workshop 2018}}},\ \bibinfo {series and number} {RWDSL2018}\ (\bibinfo  {publisher} {ACM},\ \bibinfo {year} {2018})\BibitemShut {NoStop}%
\bibitem [{Qua(2024)}]{QuantinuumH1-1}%
  \BibitemOpen
  \href@noop {} {\enquote {\bibinfo {title} {Quantinuum {H}1-1},}\ } (\bibinfo {year} {2024}),\ \Eprint {http://arxiv.org/abs/https://www.quantinuum.com/} {https://www.quantinuum.com/} \BibitemShut {NoStop}%
\bibitem [{\citenamefont {Whitfield}, \citenamefont {Biamonte},\ and\ \citenamefont {Aspuru-Guzik}(2011)}]{Whitfield2011}%
  \BibitemOpen
  \bibfield  {author} {\bibinfo {author} {\bibfnamefont {J.~D.}\ \bibnamefont {Whitfield}}, \bibinfo {author} {\bibfnamefont {J.}~\bibnamefont {Biamonte}}, \ and\ \bibinfo {author} {\bibfnamefont {A.}~\bibnamefont {Aspuru-Guzik}},\ }\bibfield  {title} {\enquote {\bibinfo {title} {Simulation of electronic structure hamiltonians using quantum computers},}\ }\href {\doibase 10.1080/00268976.2011.552441} {\bibfield  {journal} {\bibinfo  {journal} {Molecular Physics}\ }\textbf {\bibinfo {volume} {109}},\ \bibinfo {pages} {735--750} (\bibinfo {year} {2011})}\BibitemShut {NoStop}%
\bibitem [{\citenamefont {Pritchard}\ \emph {et~al.}(2019)\citenamefont {Pritchard}, \citenamefont {Altarawy}, \citenamefont {Didier}, \citenamefont {Gibsom},\ and\ \citenamefont {Windus}}]{Pritchard2019}%
  \BibitemOpen
  \bibfield  {author} {\bibinfo {author} {\bibfnamefont {B.~P.}\ \bibnamefont {Pritchard}}, \bibinfo {author} {\bibfnamefont {D.}~\bibnamefont {Altarawy}}, \bibinfo {author} {\bibfnamefont {B.}~\bibnamefont {Didier}}, \bibinfo {author} {\bibfnamefont {T.~D.}\ \bibnamefont {Gibsom}}, \ and\ \bibinfo {author} {\bibfnamefont {T.~L.}\ \bibnamefont {Windus}},\ }\bibfield  {title} {\enquote {\bibinfo {title} {A new basis set exchange: An open, up-to-date resource for the molecular sciences community},}\ }\href {\doibase 10.1021/acs.jcim.9b00725} {\bibfield  {journal} {\bibinfo  {journal} {J. Chem. Inf. Model.}\ }\textbf {\bibinfo {volume} {59}},\ \bibinfo {pages} {4814--4820} (\bibinfo {year} {2019})}\BibitemShut {NoStop}%
\bibitem [{\citenamefont {Woon}\ and\ \citenamefont {Dunning}(1994)}]{Woon1994}%
  \BibitemOpen
  \bibfield  {author} {\bibinfo {author} {\bibfnamefont {D.~E.}\ \bibnamefont {Woon}}\ and\ \bibinfo {author} {\bibfnamefont {T.~H.}\ \bibnamefont {Dunning}},\ }\bibfield  {title} {\enquote {\bibinfo {title} {Gaussian basis sets for use in correlated molecular calculations. iv. calculation of static electrical response properties},}\ }\href {\doibase 10.1063/1.466439} {\bibfield  {journal} {\bibinfo  {journal} {J. Chem. Phys.}\ }\textbf {\bibinfo {volume} {100}},\ \bibinfo {pages} {2975--2988} (\bibinfo {year} {1994})}\BibitemShut {NoStop}%
\bibitem [{\citenamefont {Neese}\ and\ \citenamefont {Valeev}(2011)}]{Neese2011}%
  \BibitemOpen
  \bibfield  {author} {\bibinfo {author} {\bibfnamefont {F.}~\bibnamefont {Neese}}\ and\ \bibinfo {author} {\bibfnamefont {E.~F.}\ \bibnamefont {Valeev}},\ }\bibfield  {title} {\enquote {\bibinfo {title} {Revisiting the atomic natural orbital approach for basis sets: Robust systematic basis sets for explicitly correlated and conventional correlated ab initio methods?}}\ }\href {\doibase 10.1021/ct100396y} {\bibfield  {journal} {\bibinfo  {journal} {J. Chem. Theory Comput.}\ }\textbf {\bibinfo {volume} {7}},\ \bibinfo {pages} {33--43} (\bibinfo {year} {2011})}\BibitemShut {NoStop}%
\bibitem [{\citenamefont {Nielsen}\ and\ \citenamefont {Chuang}(2000)}]{Nielsen2000}%
  \BibitemOpen
  \bibfield  {author} {\bibinfo {author} {\bibfnamefont {M.~A.}\ \bibnamefont {Nielsen}}\ and\ \bibinfo {author} {\bibfnamefont {I.~L.}\ \bibnamefont {Chuang}},\ }\href@noop {} {\emph {\bibinfo {title} {Quantum Computation and Quantum Information}}}\ (\bibinfo  {publisher} {Cambridge University Press},\ \bibinfo {year} {2000})\BibitemShut {NoStop}%
\bibitem [{Emu(2024)}]{EmulatorNoiseModel}%
  \BibitemOpen
  \href@noop {} {\enquote {\bibinfo {title} {{Q}uantinuum {E}mulator {N}oise {M}odel},}\ } (\bibinfo {year} {2024}),\ \Eprint {http://arxiv.org/abs/https://docs.quantinuum.com/h-series/user\_guide/emulator\_user\_guide/noise\_model.html} {https://docs.quantinuum.com/h-series/user\_guide/emulator\_user\_guide/noise\_model.html} \BibitemShut {NoStop}%
\bibitem [{\citenamefont {Makhlin}(2002)}]{Makhlin2002}%
  \BibitemOpen
  \bibfield  {author} {\bibinfo {author} {\bibfnamefont {Y.}~\bibnamefont {Makhlin}},\ }\bibfield  {title} {\enquote {\bibinfo {title} {Nonlocal properties of two-qubit gates and mixed states, and the optimization of quantum computations},}\ }\href {\doibase 10.1023/A:1022144002391} {\bibfield  {journal} {\bibinfo  {journal} {Quantum Information Processing}\ }\textbf {\bibinfo {volume} {1}},\ \bibinfo {pages} {243--252} (\bibinfo {year} {2002})}\BibitemShut {NoStop}%
\bibitem [{Sys(2024)}]{SystemModelEmulators}%
  \BibitemOpen
  \href@noop {} {\enquote {\bibinfo {title} {{Q}uantinuum {S}ystem {M}odel {H}1 {E}mulators},}\ } (\bibinfo {year} {2024}),\ \Eprint {http://arxiv.org/abs/https://docs.quantinuum.com/h-series/user\_guide/emulator\_user\_guide/emulators/h1\_emulators.html} {https://docs.quantinuum.com/h-series/user\_guide/emulator\_user\_guide/emulators/h1\_emulators.html} \BibitemShut {NoStop}%
\bibitem [{\citenamefont {Zhu}\ \emph {et~al.}(2024)\citenamefont {Zhu}, \citenamefont {Liang}, \citenamefont {Yang},\ and\ \citenamefont {Li}}]{Zhu2024}%
  \BibitemOpen
  \bibfield  {author} {\bibinfo {author} {\bibfnamefont {L.}~\bibnamefont {Zhu}}, \bibinfo {author} {\bibfnamefont {S.}~\bibnamefont {Liang}}, \bibinfo {author} {\bibfnamefont {C.}~\bibnamefont {Yang}}, \ and\ \bibinfo {author} {\bibfnamefont {X.}~\bibnamefont {Li}},\ }\href {https://arxiv.org/abs/2307.06504} {\enquote {\bibinfo {title} {Optimizing shot assignment in variational quantum eigensolver measurement},}\ } (\bibinfo {year} {2024}),\ \Eprint {http://arxiv.org/abs/2307.06504} {arXiv:2307.06504 [quant-ph]} \BibitemShut {NoStop}%
\end{thebibliography}%

\appendix

\section{Software Platforms}

A number of  software packages and platforms were used in this work.

The construction of the Hamiltonian matrix $H_{ij}$ in Fig.~\ref{fig:map}(c) was performed using the Slater-Condon rules \cite{Szabo} as implemented in Psi4NumPy \cite{Smith2018} using integrals obtained from PySCF \cite{PySCF}.

Idealised simulations used to investigate the choices of VQE hyperparameters in Fig.~\ref{fig:vqe}(c) were performed using the Qiskit Aer Simulator \cite{qiskit2024}.

Quantum circuits were constructed within Microsoft's Azure Quantum Development Kit (QDK) \cite{Microsoft_Azure_Quantum_Development} using $Q\#$ \cite{Svore_2018}.
$Q\#$ is a high-level language which allows for construction of operations across qubit registers and hybrid operations such as mid-circuit measurements, qubit re-use, integer calculations and conditional logic. 
The Azure QDK provides a hybridised workspace between Python and $Q\#$, along with connection to Azure API services and by extension connection to the Quantinuum hardware.
Quantinuum's H1-1 ion trap quantum computer \cite{QuantinuumH1-1} was the hardware of choice, along with the H1-1 emulator.
The H1-1 emulator provides an emulation target that models the hardware's ion transport and error rates, up to 20 qubits.

\section{Single-particle bases and second-quantisation}\label{app:basis}

\begin{table}
{\begin{tabular}{@{}lcccc}\toprule
Basis & N$_{\textrm{orb}}$ & $E_{\ce{He}}$ & $E_{\ce{He+}}$ & $\Delta E_{\textrm{IP}}$ (kcal/mol) \\
\colrule
cc-pVDZ		& 5  & -2.887595  & -1.993623  & 6.02 \\
cc-pVTZ		& 14 & -2.900232  & -1.998921  & 1.42 \\
cc-pVQZ		& 30 & -2.902411  & -1.999810  & 0.61 \\
cc-pV5Z		& 55 & -2.903152  & -1.999943  & 0.23 \\
\hline
ano-pVDZ & 5  & -2.897482  & -1.995072  & 0.73 \\
ano-pVTZ & 14 & -2.901703  & -1.999103  & 0.61 \\
ano-pVQZ & 30 & -2.902820  & -1.999761  & 0.32 \\
\botrule
\end{tabular}}
\caption{FCI energies from PySCF \cite{PySCF} of the \ce{He} atom and cation (in Hartree), and error in the ionisation potential (relative to the experimental value from Ref.~\onlinecite{Kandula2010}), using different commonly-used Gaussian-type basis sets.}
\label{tab:basis}
\end{table}

The most commonly used starting point for mapping electronic structure problems onto quantum computers is to use the second quantisation formalism, in which the Hamiltonian of the system is written as
\begin{equation}\nonumber 
H = \sum_{ij} t_{ij} \sum_{\sigma} a_{i\sigma}^{\dagger} a_{j\sigma} + \frac{1}{2}\sum_{ijkl} V_{ijkl} \sum_{\sigma \tau} a_{i\sigma}^{\dagger} a_{j\tau}^{\dagger} a_{l\tau} a_{k\sigma}	
\end{equation}
where $a^{\dagger}$ and $a$ are electron creation and annihilation operators, and the coefficients $t$ and $V$ are the one- and two-electron integrals obtained from a Hartree-Fock calculation.
This Hamiltonian is then written in a form suitable for implementation on a quantum computer by expanding the Hamiltonian as a linear combination of tensor products of Pauli matrices, for example using Jordan-Wigner occupation-number encoding. \cite{Whitfield2011}
With this encoding, each spin-orbital is mapped onto a unique qubit.
Then a $|1\rangle$ state of the qubit indicates the orbitals is occupied, and $|0\rangle$ indicates the orbital is unoccupied.
This second-quantised approach has the benefit that the number of terms in the Hamiltonian scales as only the fourth power of the size of the single-particle basis set employed.
However, it also has some significant drawbacks for performing high-accuracy calculations on current hardware.

To demonstrate this, Table~\ref{tab:basis} shows the errors in the IP of \ce{He} calculated using Full Configuration Interaction (FCI) on a classical computer, using different single-particle basis sets obtained from the Basis Set Exchange \cite{Pritchard2019}.
When using the widely adopted Dunning-type correlation-consistent family of basis sets \cite{Woon1994}, we see that a basis set of quadruple-zeta quality (cc-pVQZ) is necessary to obtain the IP to within 1 kcal/mol of the experimental value (chemical accuracy).
This basis set contains 30 orbitals, and mapping the resulting second-quantised Hamiltonian to the qubits of a quantum computer using the Jordan-Wigner method would require 60 qubits.
While quantum computers with more than this number of qubits have been constructed, the number of operations that would be required to produce an entangled quantum state accurately representing the ground-state of the system is beyond the capabilities of existing hardware.

The situation can be somewhat improved by using a single-particle basis set which is more appropriate for our atomic problem.
Also shown in Table~\ref{tab:basis} are the FCI results using the atomic natural orbital (ano) family of basis sets. \cite{Neese2011}
These basis sets enable chemical accuracy to be reached using as few as 5 orbitals.
However, this would still require 10 qubits using Jordan-Wigner encoding, which demonstrates the inefficiency of the second quantised approach for small problems. 
Only 25 two-electron singlet-state determinants can be constructed in the ano-pVDZ basis, but the Jordan-Wigner second-quantised encoding contains $2^{10}=1024$ computational basis states.
The approach described in Sec.~\ref{sec:rep}, which maps determinants to computational basis states, is far more qubit-efficient. All the calculations reported in this work used the ano-pVDZ basis.

\section{Hardware-adapted ansatz circuit}\label{app:circuits}

A suitable quantum circuit ansatz for our problem is shown in Fig.~\ref{fig:vqe}(a) and recreated here:

\resizebox{6cm}{!}{
\begin{quantikz} 
\lstick{\ket{0}} & \qw & \gate{R_{y}(\theta_{2})} & \qw & \qw \\
\lstick{\ket{0}} & \gate{R_y(\theta_{1})} & \ctrl{-1} & \gate{X} & \qw
\end{quantikz}
}
\\
Starting from the $|00\rangle$ (Hartree-Fock) state, this 2-parameter circuit can introduce the superposition of states required to described our wavefunction.
Given the inherent noise in quantum hardware, it is important to adapt the quantum circuits as much as possible to target the capabilities of the hardware itself.
This can involve using only gates native to the target hardware, and circuit layouts that respect as much as possible the physical connectivity between hardware qubits.
In this section we present derivations of the circuits we use on the Quantinuum H1-1 hardware and emulator.

To use the native gates of the Quantinuum H1-1 hardware, we aim to write the controlled $R_{y}(\theta)$ gate in terms of an $R_{zz}(\theta)$ gate.
To begin, we make a change of basis to a controlled $R_{z}(\theta)$ gate using the following mapping (aligning the top qubit to the right subsystem),
\begin{equation}
CR_{y}(\theta) = (I \otimes SH) CR_{z}(\theta) (I \otimes H S^{\dag})
\end{equation}
Following Ref.~\onlinecite{Nielsen2000}, the definitions of the above gates are
\begin{equation}\nonumber
I =
\begin{bmatrix}
1 & 0 \\
0 & 1 
\end{bmatrix}
\end{equation}
\begin{equation}\nonumber
H = \frac{1}{\sqrt{2}}
\begin{bmatrix}
1 & 1 \\
1 & -1 
\end{bmatrix}
\end{equation}
\begin{equation}\nonumber
S =
\begin{bmatrix}
1 & 0 \\
0 & i 
\end{bmatrix}
\end{equation}
\begin{equation}\nonumber
CR_{z}(\theta) =
\begin{bmatrix}
1 & 0 & 0 & 0 \\
0 & 1 & 0 & 0 \\
0 & 0 & e^{-i\theta/2} & 0 \\
0 & 0 & 0 & e^{i\theta/2}
\end{bmatrix}
\end{equation}
The matrix forms of the operations across both qubits are
\begin{equation}\nonumber 
I \otimes SH = 	
\begin{bmatrix}
1 & 0 \\
0 & 1 
\end{bmatrix}
\otimes
\frac{1}{\sqrt{2}}
\begin{bmatrix}
1 & 0 \\
0 & i 
\end{bmatrix}
\begin{bmatrix}
1 & 1 \\
1 & -1 
\end{bmatrix}
=
\begin{bmatrix}
1 & 1 & 0 & 0 \\
i & -i & 0 & 0 \\
0 & 0 & 1 & 1 \\
0 & 0 & i & -i	
\end{bmatrix}
\end{equation}
\begin{equation}\nonumber 
I \otimes HS^{\dag} = 	
\begin{bmatrix}
1 & 0 \\
0 & 1 
\end{bmatrix}
\otimes
\frac{1}{\sqrt{2}}
\begin{bmatrix}
1 & 1\\
1 & -1 
\end{bmatrix}
\begin{bmatrix}
1 & 0 \\
0 & -i 
\end{bmatrix}
=
\begin{bmatrix}
1 & -i & 0 & 0 \\
1 & i & 0 & 0 \\
0 & 0 & 1 & -i \\
0 & 0 & 1 & i	
\end{bmatrix}
\end{equation}
which results in
\begin{equation}\nonumber 
CR_{y}(\theta) = 
\begin{bmatrix}
1 & 0 & 0 & 0 \\
0 & 1 & 0 & 0 \\
0 & 0 & \cos(\theta/2) & -\sin(\theta/2) \\
0 & 0 & \sin(\theta/2) & \cos(\theta/2)
\end{bmatrix}	
\end{equation}
Therefore, writing the controlled $R_{y}(\theta)$ gate in terms of a controlled $R_{z}(\theta)$ gate leads to the circuit

\resizebox{8cm}{!}{
\begin{quantikz} 
\lstick{\ket{0}} & \gate{S^{\dagger}} & \gate{H} & \gate{R_{z}(\theta_{2})} &  \gate{H} & \gate{S} \\
\lstick{\ket{0}} & \gate{R_y(\theta_{1})} & \qw & \ctrl{-1} & \qw & \gate{X} & \qw
\end{quantikz}
}

The controlled-$R_{z}$ gate is not native to the Quantinuum H1-1 hardware, instead we can map to it via the $R_{zz}$ gate and the $R_{z}$ gate.
The mapping is
\begin{equation}\nonumber
(I \otimes R_{z}(\theta/2)) (R_{zz}(-\theta/2)) = 
\begin{bmatrix}
1 & 0 & 0 & 0 \\
0 & 1 & 0 & 0 \\
0 & 0 & e^{-i\theta/2} & 0 \\
0 & 0 & 0 & e^{i\theta/2}	
\end{bmatrix}
= CR_{z}(\theta)
\end{equation}
Therefore our new circuit which is built from native Quantinuum 2-qubit gates (and non-native 1-qubit gates) is:

\resizebox{8cm}{!}{
\begin{quantikz} 
\lstick{\ket{0}} & \gate{S^{\dagger}} & \gate{H} & \gate{R_{z}(\theta_{2}/2)} & \gate[wires=2]{R_{ZZ}(-\theta_{2}/2)}& \gate{H} & \gate{S} \\
\lstick{\ket{0}} & \gate{R_y(\theta_{1})} & \qw & \qw & \qw & \gate{X} & \qw
\end{quantikz}
}
\\
\\

The H1-1 emulator models the error in the $R_{zz}(\theta)$ gate using asymmetric depolarising noise.
The strength of this noise is linearly dependent on the angle of rotation $\theta$. 
More specifically it is proportional to
\begin{equation}
(przz_{a} \times \frac{|\theta|}{\pi} + przz_{b}) \times p_{2}	
\end{equation}
where $przz_{a,b} \in {\rm I\!R}^{+}$ are device-dependent fit parameters of the H1-1 hardware and $p_{2} \in [0,1]$ is the probability of a fault occuring during a $R_{zz}(\pi/2)$ gate. \cite{EmulatorNoiseModel}
Note that the $R_{zz}(\pi/2)$  gate and the conventional CNOT gate are locally equivalent. \cite{Makhlin2002}
It is assumed in this work that the model used in the H1-1 emulator for the $R_{zz}(\theta)$ gate closely matches the error relationship on the H1-1 hardware.
For the circuit and the observables measured in this work, the H1-1 emulator results closely matched the H1-1 hardware results.
At the time of use, the probability of a fault occurring in an $R_{zz}(\pi/2)$ gate was $8.8\times 10^{-4}$.
In contrast, using the arbitrary-angle ZZ gate for the optimal VQE parameters reported in Sec.~\ref{sec:vqe} results in the probability of a fault occurring during the 2-qubit gate to be only $1.85 \times 10^{-4}$.
This reduces the 2-qubit gate error to be approximately an order of magnitude lower than measurement error ($1\times 10^{-3}$ and $4\times 10^{-3}$ for both computational states respectively.) \cite{SystemModelEmulators}

To measure the bases required, additional gates are needed at the end of the circuit.
The circuits required for each basis can be simplified due to operators that do not change observables in the (measured) computational basis or by shortening single-qubit gate combinations, to further reduce the impact of noise.
It should be noted that the $S^{\dag}$ on the top qubit is also removed from all of the following as it operates on the initialised $|0\rangle$ state.
Below we show the circuits used, and simplifications made, for each measurement basis.

\textbf{YY basis}.
Top qubit: use the identity $HS^{\dag}SH = I$. 
Bottom qubit: use the identity $HS^{\dag}X = -iS\sqrt{X}^{\dag}$ and remove the $S$ gate (and global phase):

\resizebox{8cm}{!}{
\begin{quantikz}
\lstick{\ket{0}} & \gate{H} & \gate{R_{z}(\theta_{2}/2)} & \gate[wires=2]{R_{zz}(-\theta_{2}/2)} & \qw	\\
\lstick{\ket{0}} & \gate{R_{y}(\theta_{1})} & \qw & \qw & \gate{\sqrt{X}^{\dagger}}
\end{quantikz} 
}

\textbf{ZZ basis}.
Top qubit: remove the $S$ gate.
Bottom qubit: unchanged:

\resizebox{8cm}{!}{
\begin{quantikz}
\lstick{\ket{0}} & \gate{H} & \gate{R_{z}(\theta_{2}/2)} & \gate[wires=2]{R_{zz}(-\theta_{2}/2)} & \gate{H}	\\
\lstick{\ket{0}} & \gate{R_{y}(\theta_{1})} & \qw & \qw & \gate{X}
\end{quantikz} 
}

\textbf{XZ basis}.
Top qubit: remove $S$ gate.
Bottom qubit: use the identity $HX = ZH$ and remove the unobservable $Z$ gate:

\resizebox{8cm}{!}{
\begin{quantikz}
\lstick{\ket{0}} & \gate{H} & \gate{R_{z}(\theta_{2}/2)} & \gate[wires=2]{R_{zz}(-\theta_{2}/2)} & \gate{H}	\\
\lstick{\ket{0}} & \gate{R_{y}(\theta_{1})} & \qw & \qw & \gate{H}
\end{quantikz} 
}

\textbf{ZX basis}.
Top qubit: use the identity $HSH = \sqrt{X}$.
Bottom qubit: unchanged:

\resizebox{8cm}{!}{
\begin{quantikz}
\lstick{\ket{0}} & \gate{H} & \gate{R_{z}(\theta_{2}/2)} & \gate[wires=2]{R_{zz}(-\theta_{2}/2)} & \gate{\sqrt{X}}	\\
\lstick{\ket{0}} & \gate{R_{y}(\theta_{1})} & \qw & \qw & \gate{X}
\end{quantikz} 
}

\textbf{XX basis}.
Top qubit: use the identity $HSH = \sqrt{X}$.
Bottom qubit: use the identity $HX = ZH$ and remove the unobservable $Z$ gate:

\resizebox{8cm}{!}{
\begin{quantikz}
\lstick{\ket{0}} & \gate{H} & \gate{R_{z}(\theta_{2}/2)} & \gate[wires=2]{R_{zz}(-\theta_{2}/2)} & \gate{\sqrt{X}}	\\
\lstick{\ket{0}} & \gate{R_{y}(\theta_{1})} & \qw & \qw & \gate{H}
\end{quantikz} 
}

\section{Classical optimisation}\label{app:opt}

\begin{figure}
\begin{center}
\subfigure[]{
\resizebox*{8cm}{!}{\includegraphics{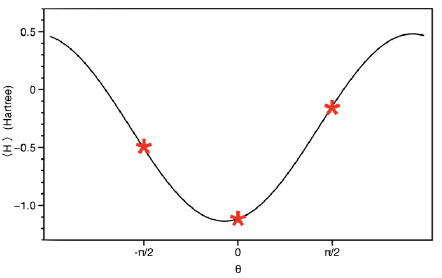}}}
\subfigure[]{
\resizebox*{8cm}{!}{\includegraphics{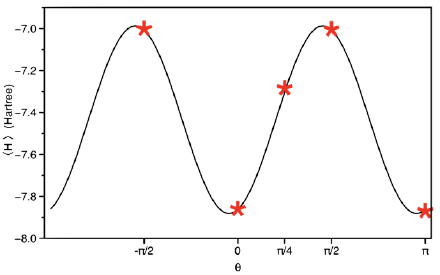}}}
\caption{\label{fig:roto}  Expectation value of a Hamiltonian as a function of the variational angle for (a): original 1-qubit-gate Rotosolve for $R_{y}(\theta_{1})$. (b): modified 2-qubit-gate Rotosolve for $R_{z}(\theta_{2})R_{zz}(-\theta_{2})$. The red stars show the special angles.}
\end{center}
\end{figure}

We use a modified version of the Rotosolve method from Ref.~\onlinecite{Ostaszewski2021}.
The original Rotosolve method provides an analytic expression for the optimal value $\theta^{*}$ of the parameter in a simple $R_{x,y,z}$ gate (holding the values of all other parameters fixed), which can be obtained by measuring the expectation value of the Hamiltonian at just three special values of the parameter (shown in Fig.~\ref{fig:roto}(a)),
\begin{eqnarray*}
\theta^{*}  =& - \frac{\pi}{2} - \arctan2 ( 2\langle H \rangle_{0} - \langle H \rangle_{\frac{\pi}{2}} - \langle H \rangle_{-\frac{\pi}{2}} , \\
& \langle H \rangle_{\frac{\pi}{2}} - \langle H \rangle_{-\frac{\pi}{2}} )
\end{eqnarray*}
where for simplicity we set the parameter $\phi$ from Ref.~\onlinecite{Ostaszewski2021} to zero.
This approach is valid for our first variational parameter $\theta_{1}$, but does not hold for the second parameter $\theta_{2}$, which appears in our circuit in both the $R_{z}$ gate and the $R_{zz}$ gate.

The expectation value of an arbitrary Hamiltonian as a function of $\theta_{2}$ (again holding all other parameters fixed) can be obtained by simple matrix multiplication using the definitions of the $R_{z}$ and $R_{zz}$ gates, and has the form
\begin{eqnarray*}
\langle H \rangle =&  A_{0} + A_{1}\cos(\theta) + A_{2}\sin(\theta) + \\
&  A_{3}\cos(2\theta) + A_{4}\sin(2\theta)
\end{eqnarray*}
The $A_{i}$ coefficients can be determined by measuring the expectation value at the five special values of $\theta$ shown in Fig.~\ref{fig:roto}(b), which gives
\begin{eqnarray*}
A_{0} &=& \frac{1}{4}\left( \langle H \rangle_{0} + \langle H \rangle_{\pi} + \langle H\rangle_{\frac{\pi}{2}} + \langle H \rangle_{-\frac{\pi}{2}} \right)	\\
A_{1} &=& \frac{1}{2} \left( \langle H \rangle_{0} - \langle H \rangle_{\pi} \right) \\
A_{2} &=& \frac{1}{2} \left( \langle H \rangle_{\frac{\pi}{2}} - \langle H \rangle_{-\frac{\pi}{2}} \right) \\
A_{3} &=& \frac{1}{4} \left( \langle H \rangle_{0} + \langle H \rangle_{\pi} - \langle H\rangle_{\frac{\pi}{2}} - \langle H \rangle_{-\frac{\pi}{2}} \right) \\
A_{4} &=& \langle H \rangle_{\frac{\pi}{4}} - A_{0} - \frac{1}{\sqrt{2}} A_{1} - \frac{1}{\sqrt{2}} A_{2}
\end{eqnarray*}
There is no analytical expression for the optimal value of $\theta_{2}$, and instead a simple 1D line minimisation is used.
In total, this optimisation method requires 8 measurements of the expectation value of the Hamiltonian per sweep.

\section{Optimal shot distribution}\label{app:shot}

\begin{figure}
\begin{center}
\resizebox*{8cm}{!}{\includegraphics{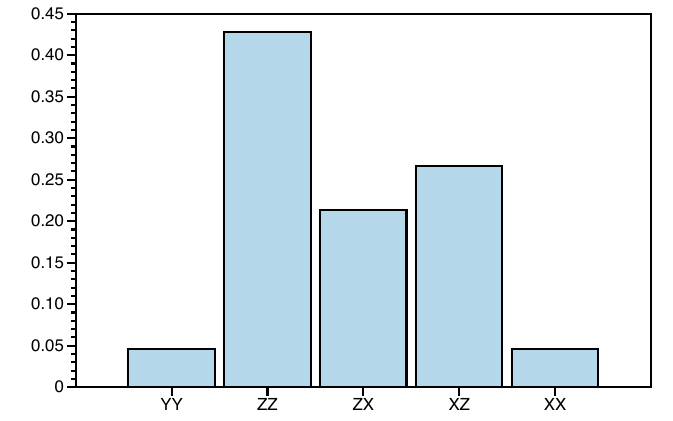}}
\caption{Optimal distribution of shots among the independent measurement bases.}
\label{fig:shots}
\end{center}
\end{figure}

To make the most of our limited hardware shot budget, we use an approach that minimises the statistical uncertainty in our estimate of the expectation value of the Hamiltonian.
This is achieved by varying the number of shots $N_{i}$ used for each of the 5 independent measurement bases (YY, ZZ, ZX, XZ, XX) while maintaining a constant total number of shots, $N$.

Following the Variance-Minimised Shot Assignment approach from Ref.~\onlinecite{Zhu2024}, we wish to minimise the variance of the measured expectation value of the Hamiltonian, which can be written in terms of the variances $\sigma_{i}^{2}$ associated with contributions from our $m=5$ measurement bases,
\begin{equation}
\min \left\lbrace \sum_{i=1}^{m} \frac{\sigma_{i}^{2}}{N_{i}} \right\rbrace, \;\; \sum_{i=1}^{m} N_{i} = N	
\end{equation}
This variance is minimised when the shots are allocated to each measurement basis in the same ratios as the standard deviations of each measurement basis,
\begin{equation}
N_{1} : N_{2} \cdots : N_{m} = \sigma_{1} : \sigma_{2} \cdots : \sigma_{m}	
\end{equation}
We estimated the values of the standard deviations $\sigma_{i}$ by performing measurements on the Quantinuum H1-1 emulator using a small number of shots, uniformly distributed across the 5 bases.
The optimal allocation of shots shown in Fig.~\ref{fig:shots} indicates that measurements of the ZZ basis dominate our estimate of the total energy, while measurements of the YY and XX bases contribute relatively little.

\end{document}